\documentclass[a4paper,12pt]{article}
\pdfoutput=1
\usepackage[utf8x]{inputenc}
\usepackage{hyperref}
\usepackage{amsfonts}
\usepackage{graphicx}
\usepackage{subfig}
\usepackage{listings}
\usepackage{amsmath,mathtools}
\usepackage{multirow}
\usepackage[linesnumbered,ruled,vlined]{algorithm2e}
\usepackage{tabularray}
\usepackage{authblk}
\usepackage[title]{appendix}
\usepackage{caption}
\title{Utilizing the International Classification of Functioning, Disability and Health (ICF) in forming a personal health index}
\author[1]{Ilkka Rautiainen\footnote{Corresponding author:\\ \textit{Email address:} \href{mailto:ilkka.t.rautiainen@jyu.fi}{\nolinkurl{ilkka.t.rautiainen@jyu.fi}}}}
\author[2]{Lauri Parviainen}
\author[2]{Veera Jakoaho}
\author[1]{Sami \"{A}yr\"{a}m\"{o}}
\author[1]{Jukka-Pekka Kauppi}
\affil[1]{Faculty of Information Technology, PO Box 35,  FI-40014 University of Jyv\"{a}skyl\"{a}, Finland}
\affil[2]{David Health Solutions Ltd., Mannerheimintie 113, FI-00280 Helsinki, Finland}
\date{}                     %% if you don't need date to appear
\providecommand{\keywords}[1]
{
	\small	
	\textbf{\textit{Keywords---}} #1
}

\begin{document}

\maketitle

\begin{abstract}

We propose a new model for comprehensively monitoring the health status of individuals by calculating a personal health index. The central framework of the model is the International Classification of Functioning, Disability and Health (ICF) developed by the World Health Organization. The model is capable of handling incomplete and heterogeneous data sets collected using different techniques. The health index was validated by comparing it to two self-assessed health measures provided by individuals undergoing rehabilitation. Results indicate that the model yields valid health index outcomes, suggesting that the proposed model is applicable in practice.

\end{abstract}


	\section{Introduction}
\label{sec:introduction}
Reliably describing and monitoring a person's health status can greatly aid in providing appropriate treatment for each individual.
Accurate information about personal health allows for better targeting of interventions, and can provide metrics that enhance factors related to overall health instead of focusing on treating a single symptom.

A personal health index aims to condense information about the overall health of a person to a single number \cite{mcdowell:_measuring_health}. While health index aims to summarize the person's health by a single number, a health profile describes the health status in a set of scores \cite{mcdowell:_measuring_health}. The index and/or profile can then be further employed in multiple ways. For instance, a singular value can provide a quick overall status of the person. This status can be useful for healthcare professionals and for the persons themselves, aiming at systematic and understandable way of monitoring the health status of persons from a broad perspective \cite{mcdowell:_measuring_health}. In addition, examining further different aspects of well-being provided by a health profile is often crucial to obtain more detailed perspective of person's health.

Concisely describing a person's overall health status can be challenging. Changes in a person's functioning and health can be attributed to several reasons. For instance, the prevention, treatment and rehabilitation of physical illness, such as back pain, include also the consideration of psychological and social aspects \cite{fava2000psychosomatic}. Therefore, it is necessary to describe all the aspects affecting a person's health in a structured and standardized form, and to present this information in an easily interpretable fashion.

Global standardization of health status measurement procedures for health index construction is difficult to achieve due to different standards and practices in different countries. There can be significant variations in treatment procedures depending on the practices of different countries, clinics and therapists. For instance, clinics in different countries may use different health questionnaires (and different languages). These differences make straightforward concatenation of data sets from separate clinics impossible.

In this study, we propose using the International Classification of Functioning, Disability and Health (ICF) \cite{who:_icf} developed by the World Health Organization as a basis for construction of a personal health index.  As a comprehensive and standardized classification of functioning, disability and health, ICF covers all relevant aspects affecting personal health, making it an ideal platform for health index development. By using the ICF framework, we first convert original measurements to new variables in “ICF space” using accepted linking procedures. These variables are called ICF codes throughout the paper according to ICF terminology. This allows standardization of possibly heterogeneous data sets from different individuals into the same data space. Then, we recursively calculate the health index from an ICF tree structure using available measurements. Calculation does not require measurements from every node of the ICF tree, making it robust to missing values. On the other hand, if measurements do not cover all relevant aspects of health, the missing parts can be taken into account in future data collection. Once the overall health index value is calculated, it is also possible to investigate values for each ICF code in the tree separately, providing insights into specific sectors of functioning, disabilities and contextual factors affecting health. This way, it is possible to obtain a comprehensive picture of individual's health and see whether specific aspects of health need to be improved. To the best of our knowledge, this is the first study that employs the ICF framework to form a personal health index.

Multiple health indices and health profiles have been recently suggested in literature.
Meijer et al. (2010) \cite{meijer:_internationally_comparable_health_indices} introduced an internationally comparable health index that is based on functional limitations and self-reported health measures in addition to objectively measured grip strength. Kohn (2012) \cite{kohn:_what_is_health} employed multiple correspondence analysis to form a health index, giving some freedom of choice 
to the index user, such as the decision of what questions to include in the index domain. Poterba et al. (2013) \cite{poterba:_health_education_and_the_postretirement_evolution} used 
principal component analysis for the responses to 27 health-related questions, using the first principal component as their health index. 
Chen et al. (2016) \cite{chen:_personal_health_indexing_based_on_medical_examinations} developed a method called MyPHI that gives a personal health index (PHI) as its output. 
They treated the task of creating the PHI as a soft-label optimization problem with a data mining technique originally designed for complex event detection on video material. 
Their method can handle geriatric data with infrequencies, incompleteness and sparsity. It is also designed to give higher weight to the latest health records. 
Since the output of PHI is a vector of scores with each score reflecting personal health risk in a disease category, by the McDowell (2006) \cite{mcdowell:_measuring_health} definition, their method outputs a health profile instead of a health index. 
Lai et al. (2020) \cite{lai:_personal_health_index} introduced a personal health index based on a tensor decomposition method to overcome the limitations of health examination records.

Overall, in the existing research there are several elements of what we aim to achieve in this study, such as considering the differences in data by country as well as 
the ability to handle sparse and infrequent data with missing values. However, previously suggested health indices and profiles for the most part rely on a predefined set of attributes that are used to calculate the final health index. Since there are many existing methods for collecting data, for example through functioning questionnaires \cite{fairbank:_the_oswestry_low_back_pain_disability_questionnaire,herdman:_development_and_preliminary_testing} and customized tests (e.g. with equipment available at the clinic), it would be more beneficial to combine the information from different sources effectively, and then utilize it further in the health index calculation process.

The remaining part of the study proceeds as follows. Section~\ref{sec:ICF} gives a brief introduction to the ICF. This is necessary preliminary information for understanding health index calculation. Section~\ref{sec:data} describes data used in this study, including necessary conversion steps of the original data to the ICF space. Section~\ref{sec:methods} describes actual computation of a health index. Section~\ref{sec:results} presents the results produced during the statistical analyses. The study is discussed and concluded in Section~\ref{sec:discussion}.

\section{International Classification of Functioning, Disability and Health (ICF)}
\label{sec:ICF}

The overall aim of the International Classification of Functioning, Disability and Health (ICF) is to 
''provide a unified and standard language and framework for the description of health and health-related states''.
It is a complementary classification to the ICD-10/11 diagnosis classification \cite{who:_icf}.

The ICF brings together two conceptual paradigms of disability: the medical paradigm and the social paradigm. The medical paradigm sees disability as something which requires medical care, the cause being a disease, trauma or other health condition of the individual. Moreover, in the social paradigm disability is seen as a socially-created problem, something that requires a political response. The fusion of these two paradigms in the ICF can be seen as a ''biopsychosocial'' approach. In other words, the ICF provides a synthesis of biological, individual and social aspects of health \cite{who:_icf, world2002towards}.
The ICF model has potential to change the current disease-based model of care, from anticipating or reacting to individual diseases to \textit{healthy ageing} approach, which aims to observe individual's trajectories longitudinally in order to support personalized interventions proactively \cite{cesari2018evidence}.

\begin{figure}[h!]
	\centering
	\includegraphics[height=9.0cm,keepaspectratio,clip,trim=0.7cm 0.7cm 0.7cm 0.6cm]{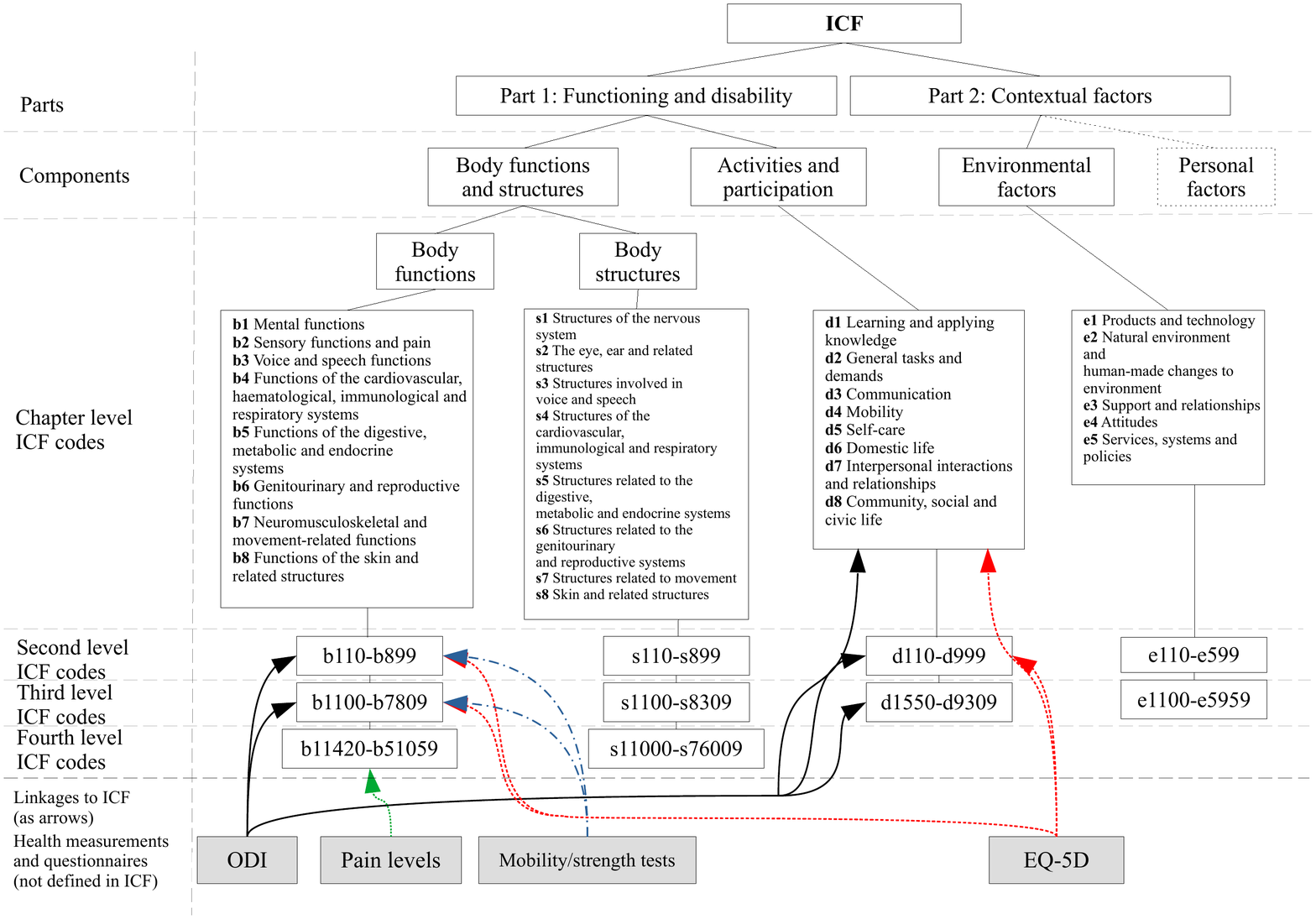}
	\caption{The hierachical structure of the ICF (combined from the information available in \cite{who:_icf}) with the utilized health measurements and questionnaires and their linkages to ICF.}
	\label{fig:ICF}
\end{figure}

An overview of the structure of the ICF is presented in Fig.~\ref{fig:ICF}. The four health measurements/questionnaires we are utilizing and linking to the ICF are shown at the bottom of the figure, and they are not part of the ICF itself. These elements are discussed later in Section~\ref{subsec:linkage_of_items}.

The ICF classifies health and health-related issues and it consists of two main parts, \textit{Functioning and disability} and \textit{Contextual factors}, which are further divided into four main \textit{components}: \textit{Body functions} ($b$) (physiological functions of body systems, including psychological functions), \textit{Body structures} ($s$) (anatomical parts of the body), \textit{Activities and participation} ($d$) (execution of a task or action by an individual and involvement in a life situation) and \textit{Environmental factors} ($e$) (physical, social and attitudinal environment in which people live). Although \textit{Personal factors} (background of an individual's life and living) are part of the structure, they can not currently be classified using ICF \cite{who:_icf}.

The main components have over 1,400 subcategories that are spread over four hierarchical levels. The categories in ICF are nested; the broader categories are defined to include more detailed subcategories of the parent category. For example, the chapter two ICF code, \textit{The eye, ear and related structures} ($s2$), in the \textit{Body structures} component includes separate categories on the structures of eye socket, eyeball, around eye and so on \cite{who:_icf}.

The \textit{chapter level} ICF code is indicated by writing a single digit number after the component $b$, $s$, $d$ or $e$, for example $b2$ referring to the ICF code \textit{Sensory functions and pain}. The \textit{second level} ICF code is specified using a three digit number (e.g. $b280$ is the \textit{Sensation of pain}). Furthermore, the \textit{third level} ICF code has four numbers (e.g. $b2801$ is the \textit{Pain in body part}), and finally the \textit{fourth level} has five numbers (e.g. $b28013$ is the \textit{Pain in back}) \cite{who:_icf, world2002towards}.

To become a \textit{classification} and therefore complete, the ICF code is equipped with one or more \textit{qualifiers}. A qualifier is a single-digit number that is used to denote, for instance, the magnitude of the level of health or severity of the problem at issue. When the qualifier is written for an ICF code, a dot ($.$) is used as a separator between the ICF code and the qualifier(s). For example, a complete ICF code with a qualifier would be $b280.1$, which would indicate that the person is feeling a mild/slight \textit{Sensation of pain} \cite{who:_icf}.

All of the possible ICF codes support the first qualifier, though its usage varies slightly between different main constructs. For the ICF codes under \textit{Body functions} ($b$) and \textit{Body structures} ($s$) the first qualifier is a \textit{generic qualifier} that indicates the extent or magnitude of an impairment. A generic qualifier can have values 0 (\textbf{no} problem), 1 (\textbf{mild} problem), 2 (\textbf{moderate} problem), 3 (\textbf{severe} problem) or 4 (\textbf{complete} problem), with additional values 8 and 9 reserved for not specified and not applicable states, respectively. For the ICF codes under the \textit{Activities and participation} ($d$) component the first generic qualifier indicates \textit{performance}, a problem in the person's current environment \cite{who:_icf, world2002towards}.

In the case of \textit{Environmental factors} ($e$), the generic qualifier can be either a \textit{barrier} or a \textit{facilitator}, meaning that a positive contributor can also be recorded. The negative scale is employed similarly to the aforementioned constructs for the barriers, while the facilitators are indicated using a plus sign ($+$) as an ICF code/qualifier separator instead of a dot (e.g. ICF code $e145+2$ would indicate that products for education are a moderate facilitator, as opposed to $e145.2$, which would mean that the products of education are a moderate barrier) \cite{who:_icf, world2002towards}. The facilitators are not, however, part of our proposed health index calculation model.

In addition to the first qualifier, it is possible to define additional qualifiers for \textit{Body structures} and \textit{Activities and participation}. For example, the nature of impairment can be described using the second qualifier in \textit{Body structures} (e.g. $s7300.32$ indicates a partial absence of the upper extremity) \cite{who:_icf, world2002towards}. However, since we utilize only the first qualifier in our proposed calculation model, these additional qualifiers are not discussed here further. Throughout the rest of this paper, the term \textit{qualifier} will refer only to the first qualifier.

\section{Data}
\label{sec:data}

For the development and validation of the health index, we employed data collected from a single David Health Solutions clinic between 2013--2019. The data included 505 persons (age $48\pm18$ y, min 12 y, max 87 y, 259 females and 246 males) receiving rehabilitation treatment for various problems, including back, neck, hip, knee, shoulder, general health, and other unspecified reasons. Our original data consisted of following questionnaire and measurement data sets:

\begin{itemize}
	\item Oswestry low back pain disability questionnaire (ODI) \cite{fairbank:_the_oswestry_low_back_pain_disability_questionnaire}
	\item a generic health questionnaire EQ-5D-5L \cite{herdman:_development_and_preliminary_testing}
	\item mobility and maximal isometric strength tests using various spine concept rehabilitation machines
	\item pain level answers
\end{itemize}

A required preprocessing step in the health index construction is to convert, or link, variables of these original data sets to new variables called ICF codes. Linkage of original variables to ICF codes is described next.

\subsection{Linkage of data to the ICF}
\label{subsec:linkage_of_items}

The ICF linkage guidelines \cite{cieza2002,cieza2019} recommend that two medical professionals are trained to create the linkages between the original data sources to the ICF code qualifiers. In the linking process, two experts independently created the linkage information for a new data source. If the two independent linkages were identical they were accepted as is. In case they differed, a third expert opinion was considered when deciding the final linkage. Each measurement $s,m \in \mathbb{Z}$ was always mapped linearly to the range $0 \leq x \leq 4$, which is the range of the qualifiers in the ICF.
Details of linkage for each questionnaire and measurement data set are available in \ref{app:data_linkage_tables}.

Since there can be both 
scientifically validated and self-made ICF linkages present in the data, it is essential to be able to define separate linkage reliability values for different linkage types. $r \in [0,1]$ is the \textbf{linkage reliability} defined for the corresponding source. The index user must define the $r$ value for every available source before calculating the index. For instance, we might want to define independently validated linkages with maximum reliability 1, while self-made linkages can be considered less reliable. Thus, their $r$ would in most cases be defined to be less than 1.

For the two questionnaires, EQ-5D and ODI, data were available for 111 and 147 persons, respectively. The pain level answers were accessible for 348 persons and mobility/strength test results for 420 persons. Additionally, EQ-VAS, self-assessed health status, was answered by 168 persons.

\begin{figure}[h!]
	\centering
	\includegraphics[height=5cm,keepaspectratio]{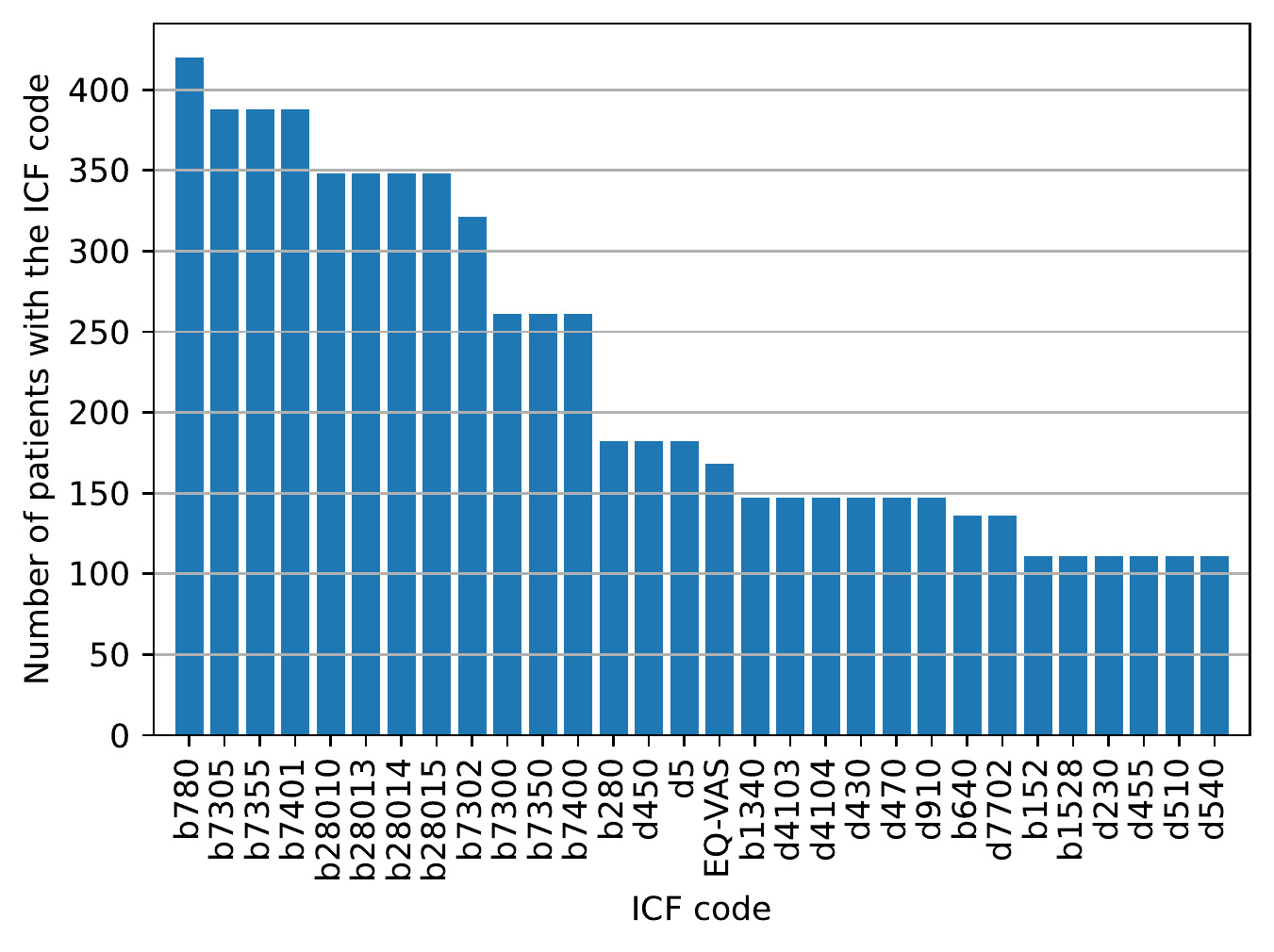}
	\caption{An overview of the occurrence of ICF codes in the David dataset for the persons. Additionally, EQ-VAS answer is also included.}
	\label{fig:ICF_codes_vs_patients}
\end{figure}

Most often available ICF code in the dataset was $b780$ (\textit{Sensations related to muscles and movement functions}), which was available for 420 persons out of 505. The next three accessible ICF codes were $b7305$ (\textit{Power of muscles of the trunk}), $b7355$ (\textit{Tone of muscles of trunk}), and $b7401$ (\textit{Endurance of muscle groups}). They were all available for 388 persons. See Fig.~\ref{fig:ICF_codes_vs_patients} for a full list of available ICF codes. To be included in the list, there had to be at least one measurement available for the specified ICF code.

\subsection{Overview of data}\label{subsec:overview_of_data}

There are two important properties to examine in the data. Firstly, \textit{duration} of a treatment period is the number of full calendar days between the beginning and the end of the latest day of the treatment. By this definition, when treatment is made only once, the duration of the treatment is 0. Secondly, we can examine the frequency or number of treatment days during the treatment period. We call this property a \textit{length} of the treatment sequence. For example, if we have a person that has been treated for three weeks and had four visits to the clinic, the duration of the treatment period would be 21 days, with four as the length of the treatment sequence.

\begin{figure}[ht!]
	\centering
	\subfloat[]{\includegraphics[width=6.4cm]{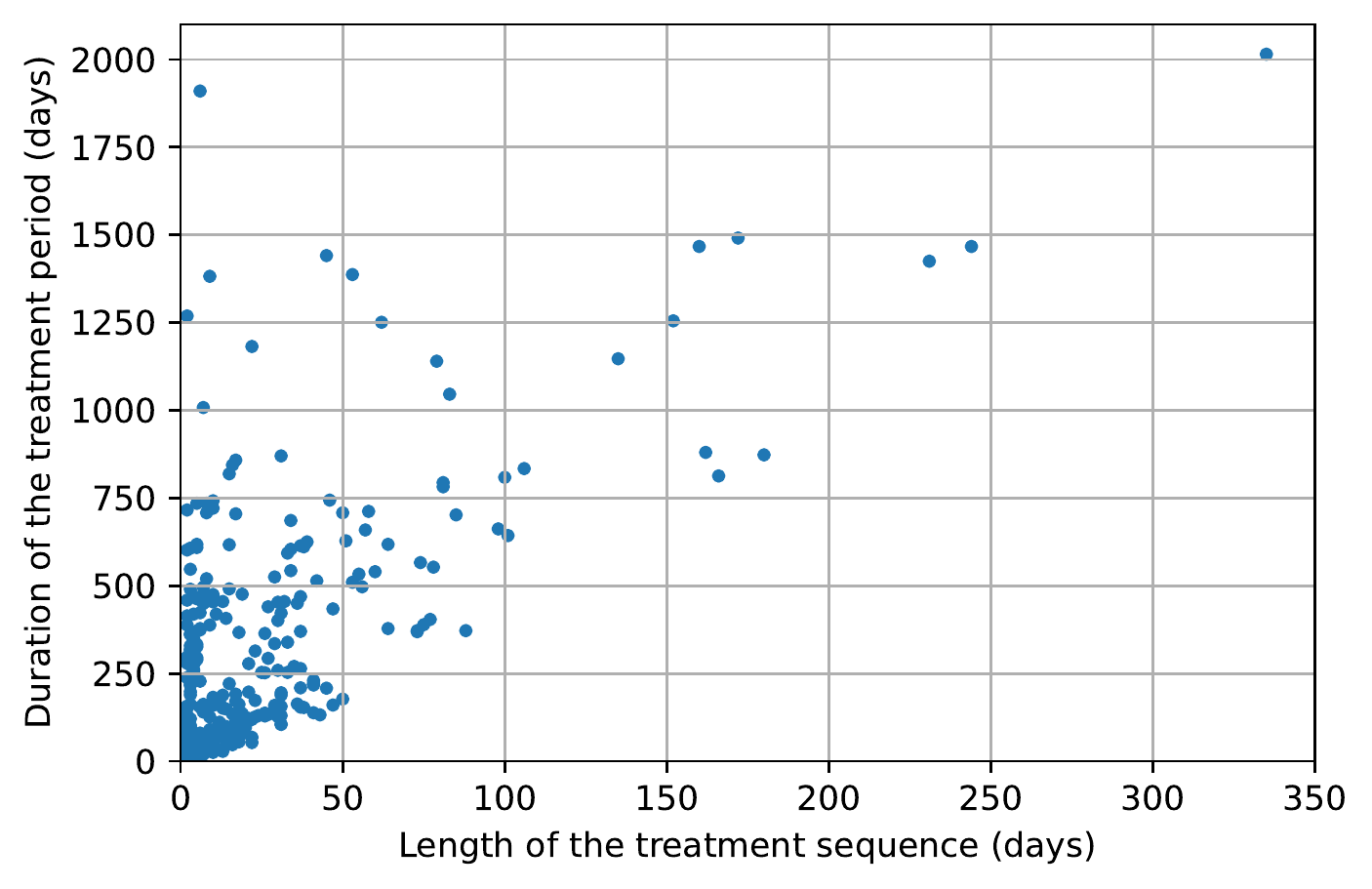}\label{fig:no_of_measurement_days_vs_length_of_treatment_a}}%
	\qquad
	\subfloat[]{\includegraphics[width=6.4cm]{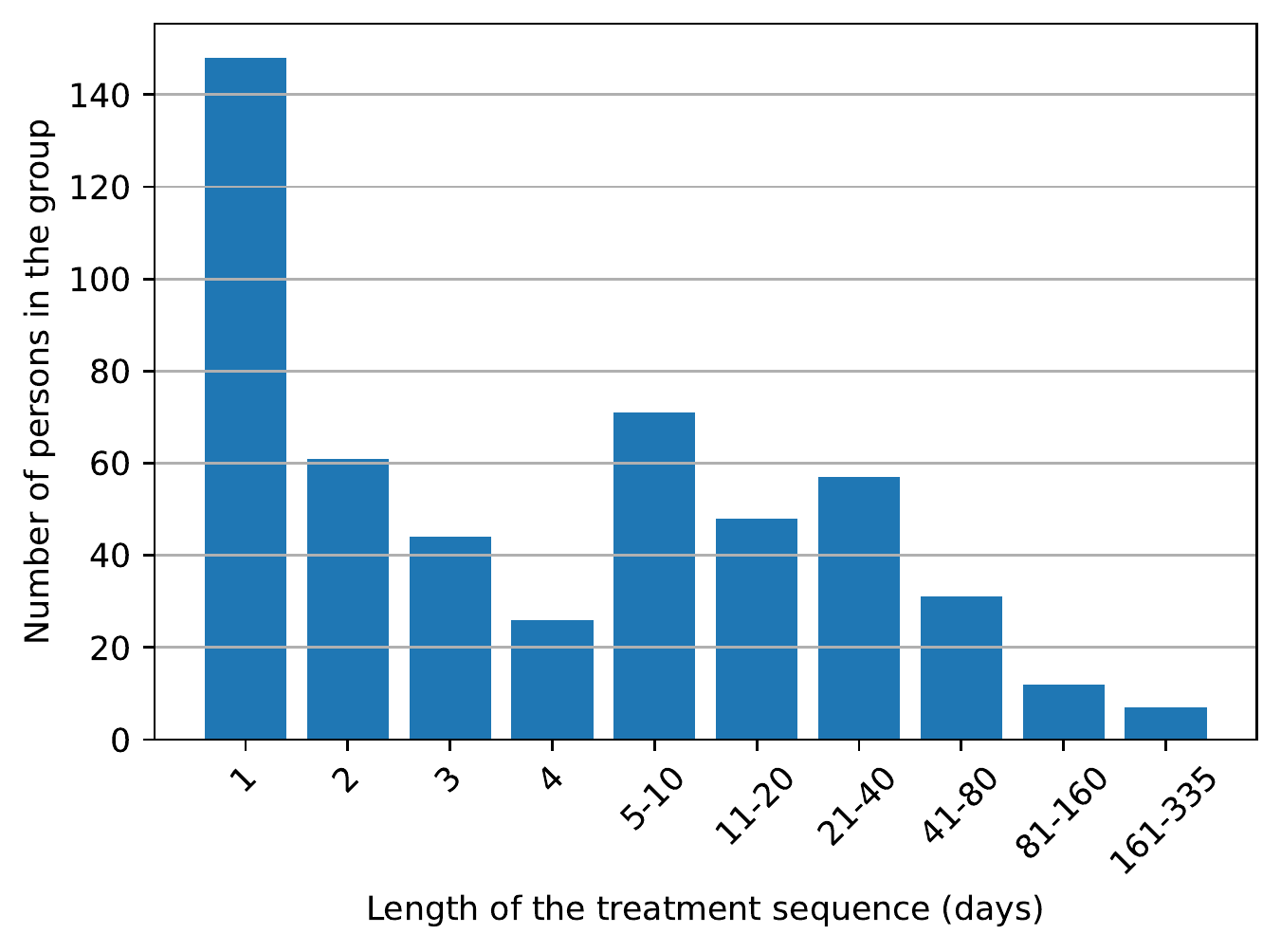}\label{fig:no_of_measurement_days_vs_length_of_treatment_b}}%
	\caption{In \protect\subref{fig:no_of_measurement_days_vs_length_of_treatment_a} duration of the treatment periods are shown. The duration of the treatment period and the length of the treatment sequence are presented for each person. The majority of the persons are in the lower-left corner, meaning that they have had relatively short treatment with few individual treatment days. \protect\subref{fig:no_of_measurement_days_vs_length_of_treatment_b} shows how many days there is data from. Ten separate groups for the length of the treatment sequences were formed to examine how many persons were in each group. For instance, we see that for over 140 persons data is available for one day only.}%
	\label{fig:no_of_measurement_days_vs_length_of_treatment}%
\end{figure}

The median duration of the treatment period in data was 69 days, while the median length of the treatment sequence was three days. %period included measurements from three (median) separate treatment days.
In Fig.~\ref{fig:no_of_measurement_days_vs_length_of_treatment} we examine these two properties further.

\section{Calculating the health index}
\label{sec:methods}

In this section, methods for calculating the health index are presented. We first introduce a structure of the model used in calculation, involving its basic elements. After that, we describe overall calculation process of the health index from the model. Then, we present model elements in more detail, followed by details of the ICF code level calculations.

\subsection{Model structure and elements}

The proposed model structure corresponds to the hierarchical structure of the ICF. However, instead of using all possible ICF codes, the proposed model structure consists of available ICF codes only. For instance, ICF codes available in this study are listed in Fig.~\ref{fig:ICF_codes_vs_patients}, so the structure of the model consists of these ICF codes only. Within this restricted set of all possible ICF codes, some ICF codes can further be missing, i.e. they are not measured for all persons. Thus, some ICF codes in the model are observed codes whereas some codes are empty codes. Moreover, an ICF code in the model can be observed through not only one but multiple measurements. This situation happens when more than one original data variables are linked to a same ICF code. It is also possible to have multiple measurements in a single ICF code by linking one original data variable recorded from different time points. In this study, all linked measurements are called qualifiers according to the ICF terminology.

\begin{figure}[h!]
	\centering
	\includegraphics[height=4.4cm,keepaspectratio]{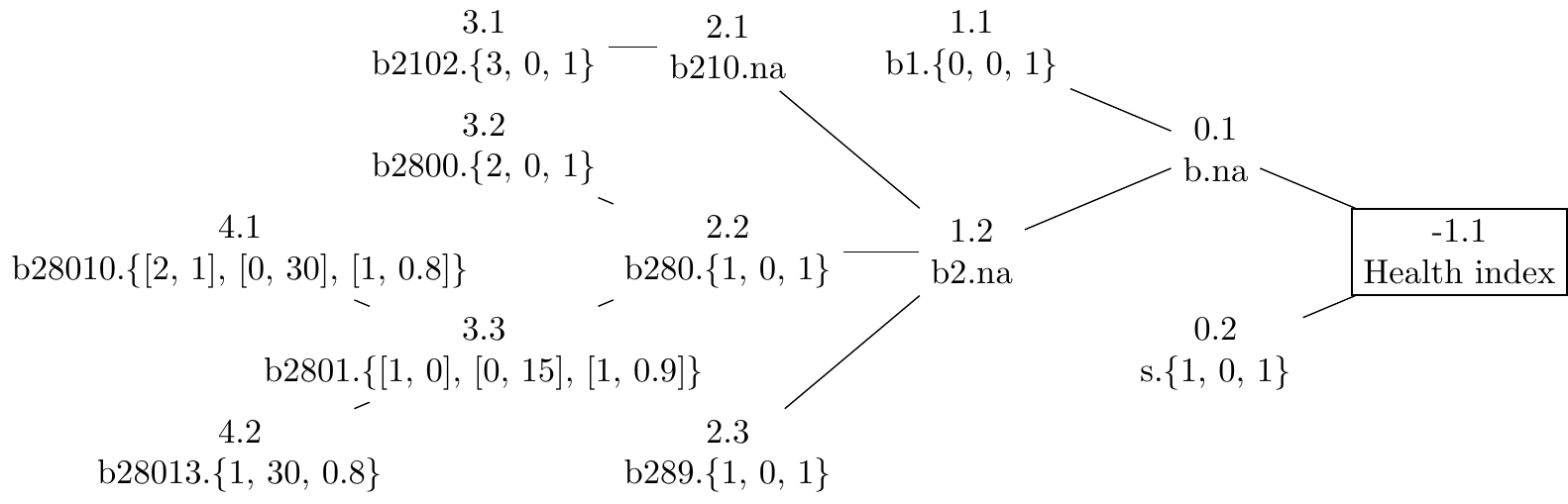}
	\caption{An example of a tree containing ICF codes. Each ICF code has information in two rows. The first row has the level number and an ICF code number in the level, separated by a dot. The format used is $level.index$, where $level$ is the ICF level the node belongs to, level 4 being the deepest, and $index$ is the per-level running index. The ICF codes are in alphabetical order at each level, and the running index is determined by this order. The level labeled as $0$ refers to the ICF top components $b$, $d$, $e$ and $s$, the level 1 to the chapter level ICF codes, the level 2 to the second level ICF codes and so on. The second ICF code row indicates the ICF code and three attributes associated with it, ICF qualifier $x$, time elapsed from measurement ($TE$) (see \ref{app:time_weighting}) and reliability of the linkage $r$ (see Section~\ref{subsec:linkage_of_items}). The format is \textit{ICF\_code.\{$qualifier$, $time\_elapsed$, $reliability$\}.} When a qualifier is not available for the ICF code, \textit{na} is shown instead after the dot.}
	\label{fig:tree_example}
\end{figure}

For illustration purposes, Fig.~\ref{fig:tree_example} gives an example of a simplified tree structure containing ICF codes.
As mentioned above, multiple original measurements can be linked to a same ICF code, meaning that the corresponding ICF code is observed through multiple qualifiers. In Fig.~\ref{fig:tree_example} we have two ICF codes with two qualifiers linked to the same ICF code on the third and fourth levels. Firstly, the first qualifier in the ICF code $b28010$ (4.1) is 2 (first number inside the first square brackets), with $time\_elapsed=0$ and $reliability=1$. The second qualifier, a measurement coming from a different source, for the same ICF code is 1 (second number inside the first square brackets), with $time\_elapsed=30$ and $reliability=0.8$. Finally, the second ICF code with two qualifiers is $b2801$ (3.3). Also empty ICF codes are included in Fig.~\ref{fig:tree_example} for illustration purposes.

\subsection{Health index calculation}
\label{subsec:health_index_calculation}

The principal idea in the proposed approach is to transport information node-by-node and level-by-level from the deepest level to the root of the tree, 
eventually resulting in a single number called the health index. 
The calculation procedure for the health index is started from the leaf nodes of the tree. For example, in Fig.~\ref{fig:tree_example} the calculation starts from ICF code b28010 (4.1), then continuing on to b28013 (4.2) and then to the next level ICF code b2102 (3.1) etc. The calculation procedure is repeated through the tree to the right until there are no ICF codes left. The last calculated value is the unscaled health index, and scaling this value to 0--100 produces the final health index. The calculation procedure is described in Alg.~\ref{alg:hi}. The equations referred to in the algorithm are presented after the algorithm.

%algorithm2e style
\begin{algorithm}
	\small
	\DontPrintSemicolon
	\SetNoFillComment
	\KwData{ICF code tree with qualifiers}
	\KwResult{Health index}
	initialize variable \textit{level} with the deepest ICF code level in the tree\; \label{alg1:line:initialize_variable_level}
	\While{the health index has not been calculated}{\label{alg1:line:while_outer_loop}
		\textit{index} = 1\; \label{alg1:line:index}
		\While{there are ICF codes at the level}{\label{alg1:line:while_inner_loop}
			\If{there are child ICF codes for the ICF code level.index}{\label{alg1:line:if_there_are_child_ICF_codes}
				calculate value of the ICF code $x_{level.index}$ (Eqs.~\ref{eq:qx}, \ref{eq:beta_m}, \ref{eq:beta_s})\; \label{alg1:line:calculate_x}
				calculate time weighting $\alpha_{level.index}$ (Eq.~\ref{eq:weighted_mean_alpha_and_r})\; \label{alg1:line:calculate_alpha}
				calculate linkage reliability $r_{level.index}$ (Eq.~\ref{eq:weighted_mean_alpha_and_r})\; \label{alg1:line:calculate_r}
				\ForEach{child ICF code of level.index}{\label{alg1:line:foreach_child_ICF_code}
					remove all qualifiers $s$ if they exist\; \label{alg1:line:remove_all_qualifiers}
				}
				\textit{index} = \textit{index} + 1\; \label{alg1:line:index_plus}
			}
		}
		\textit{level} = \textit{level} - 1\; \label{alg1:line:level_minus}
		\If{$level < -1$}{\label{alg1:line:if_level}
			\textit{scaled\_health\_index} = $HI(x_{-1.1})$ (Eq.~\ref{eq:transformation})\; \label{alg1:line:scale_health_index}
			return \textit{scaled\_health\_index}\; \label{alg1:line:return_health_index}
		}
	}
	\caption{Health index calculation.}\label{alg:hi}
\end{algorithm}

In Alg.~\ref{alg:hi} the input data is the ICF code tree with qualifiers, similar to the one depicted in Fig.~\ref{fig:tree_example}. The output result is the health index. In the beginning (line number \ref{alg1:line:initialize_variable_level}), we initialize the variable \textit{level}, according to the deepest ICF code level available in the data. For example, in the data depicted in Fig.~\ref{fig:tree_example}, there are qualifiers present in two ICF codes at level 4 (4.1 and 4.2). Thus, the \textit{level} variable in this case would be initialized as 4. The lowest possible $level$ is $-1$, corresponding to the root of the tree. At this level, the final health index is calculated.

The purpose of the outer \textbf{while} loop (Alg.~\ref{alg:hi}, lines \ref{alg1:line:while_outer_loop}--\ref{alg1:line:return_health_index}) is to process ICF codes of the tree one by one until the root of the tree has been reached and the health index calculated and returned. Inside the loop, the variable \textit{index} is first initialized (line \ref{alg1:line:index}). The \textit{index} acts as a per-level running index, i.e. it is reset to 1 every time the level of the tree changes. For example, the processing order of the ICF codes in the tree shown in Fig.~\ref{fig:tree_example} is 4.1, 4.2, 3.1, 3.2, 3.3, and so on.

The inner \textbf{while} loop (Alg.~\ref{alg:hi}, lines \ref{alg1:line:while_inner_loop}--\ref{alg1:line:index_plus}) serves a purpose of repeating the necessary calculations until all the ICF codes in the current $level$ are handled. The first \textbf{if} statement (line \ref{alg1:line:if_there_are_child_ICF_codes}) checks whether there are child ICF codes available for the currently examined ICF code, denoted as $level.index$. For example, in Fig.~\ref{fig:tree_example} the child ICF codes of $b280$ are $b2800$ and $b2801$.

In case child ICF codes are available for an ICF code, calculations can start. In lines \ref{alg1:line:calculate_x}--\ref{alg1:line:calculate_r} the value of the ICF code, $x_{level.index}$, time weighting $\alpha_{level.index}$ and linkage reliability $r_{level.index}$ are calculated.
These calculations are discussed in detail in Sec. \ref{subsec:calculation_of_ICF_code_and_r_values}.

In lines \ref{alg1:line:foreach_child_ICF_code}--\ref{alg1:line:remove_all_qualifiers} every qualifier is removed from the child ICF codes of \textit{level.index}. The purpose of this procedure is to make sure these values are not used later in the calculations. Instead, the values calculated during lines \ref{alg1:line:calculate_x}--\ref{alg1:line:calculate_r} are employed when the calculation continues further.

The per-level running $index$ is increased after each ICF code has been handled (Alg.~\ref{alg:hi}, line \ref{alg1:line:index_plus}), and the level index $level$ decreased after all ICF codes in the level have been handled (line \ref{alg1:line:level_minus}).

When the level index $level$ goes below $-1$, it means that the root of the tree has been reached (Alg.~\ref{alg:hi}, line \ref{alg1:line:if_level}). The raw health index, denoted $x_{-1.1}$ by the ICF code numbering scheme employed in Fig.~\ref{fig:tree_example}, has been calculated. The raw value will be finally transformed to make it more decipherable. 
A suitable target presentation for the final health index is an integer number from 0 (worst health) to 100 (best health). To achieve this, the source and target values are inverted: while raw health index value 0 indicates best possible health, the same value in the target range indicates worst possible health. 

The transformation is applied in the following way, with the final health index denoted as $HI$ (Alg.~\ref{alg:hi}, line \ref{alg1:line:scale_health_index}):

\begin{equation}\label{eq:transformation}
	HI(x_{-1.1})=\textup{nint}\left ( 100-100 \times \frac{x_{-1.1}-\textup{min}(x_{-1.1})}{\textup{max}(x_{-1.1})-\textup{min}(x_{-1.1})}  \right ),
\end{equation}

where $\textup{nint}$ is the nearest integer function and $x_{-1.1}$ is the raw value to be transformed, 
while $\textup{min}(x_{-1.1})$ and $\textup{max}(x_{-1.1})$ are the minimum and maximum theoretical or actual raw values.
Since we know the theoretical minimum (0) and maximum (4) values of the raw health index, our first option is to directly scale and invert the raw values to our target. 
In this case we would define that $\textup{min}(x_{-1.1})=0$ and $\textup{max}(x_{-1.1})=4$.
The returned value after the transformation (Alg.~\ref{alg:hi}, line \ref{alg1:line:return_health_index}) is the final health index.

After the health index has been calculated, the tree is reset to its former state, meaning that all calculated values are removed from the ICF codes, and all qualifiers are restored to their original positions. This procedure's meaning is to ensure that any future calculations do not use any previously calculated values.

\subsection{A closer look at the elements}
\label{subsec:closer_look}

\begin{figure}[h!]
	\centering
	\fbox{\includegraphics[height=6.0cm,keepaspectratio,clip,trim=1.5cm 4.5cm 2cm 4.5cm]{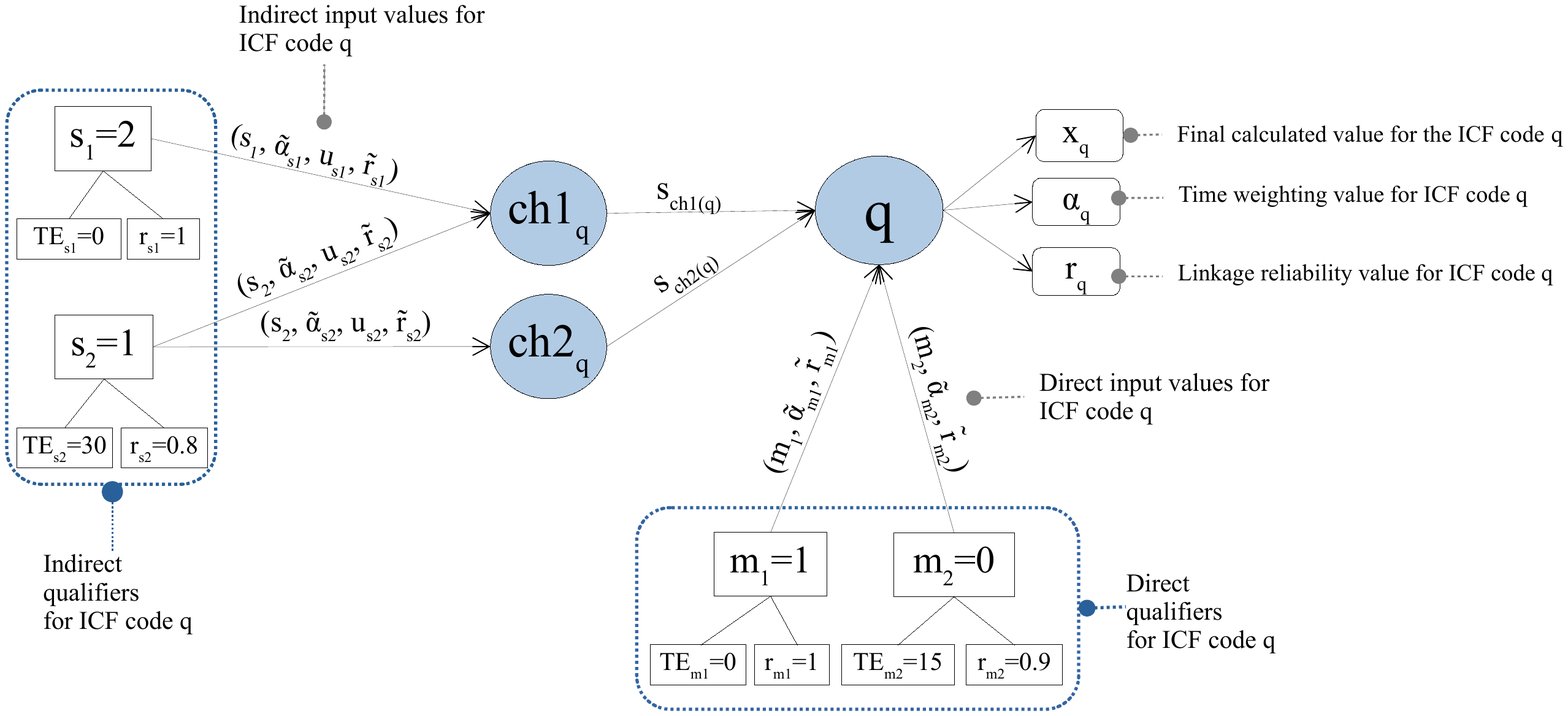}}
	\caption{An ICF code level example of the health index calculation. In this figure we have a viewpoint of an individual ICF code marked as $q$. Its child ICF codes are marked here as $ch1_{q}$ and $ch2_{q}$.}
	\label{fig:hi_calculation_diagram}
\end{figure}

In Fig.~\ref{fig:hi_calculation_diagram} we take a closer look at the elements in the tree. Here we have taken three ICF codes, $b28010$, $b28013$ and $b2801$, from the Fig.~\ref{fig:tree_example} for a more thorough inspection. An ICF code, whose output value is calculated, is denoted by $q$. In Fig.~\ref{fig:hi_calculation_diagram}, this ICF code is equivalent to node 3.3 (\textit{b2801}) shown in Fig.~\ref{fig:tree_example}. Furthermore, $q$'s child ICF codes, $ch1_{q}$ and $ch2_{q}$, are equivalent to ICF codes 4.1 (\textit{b28010}) and 4.2 (\textit{b28013}), respectively.

In this hypothetical sample data there are four separate qualifiers, indicated with markings $s_1$, $s_2$, $m_1$ and $m_2$, with linkages between qualifiers and ICF codes indicated with arrows:

\begin{enumerate}
	\item A \textit{moderate problem} (qualifier $s_1=2$) is recorded for ICF code \textit{Pain in head and neck} ($b28010$/$ch1_{q}$). The measurement is made in the latest available day in the treatment period, as indicated by $TE_{s1}=0$ and its linkage reliability is defined as the strongest possible ($r_{s1}=1$).
	\item A \textit{slight problem} (qualifier $s_2=1$) is recorded for ICF codes \textit{Pain in head and neck} ($b28010$/$ch2_{q}$) and \textit{Pain in back} ($b28013$/$ch2_{q}$). The measurement is made 30 days ago from the perspective of latest available measurement, as indicated by $TE_{s2}=30$, and its linkage reliability is defined as ($r_{s2}=0.8$).
	\item A \textit{slight problem} (qualifier $m_1=1$) is recorded for ICF code \textit{Pain in body part} ($b2801$/$q$). The measurement is made in the latest available day in the treatment period, as indicated by $TE_{m1}=0$ and its linkage reliability is defined as the strongest possible ($r_{m1}=1$).
	\item A \textit{no problem} (qualifier $m_2=0$) is recorded for ICF code \textit{Pain in body part} ($b2801$/$q$). The measurement is made 15 days ago from the perspective of latest available measurement, as indicated by $TE_{m2}=15$, and its linkage reliability is defined as ($r_{m2}=0.9$).
\end{enumerate}

The different elements are marked in Fig.~\ref{fig:hi_calculation_diagram}:

\begin{itemize}
	\item \textbf{Direct qualifiers (D) for ICF code q:} $m_{1}$ and $m_{2}$ are the \textbf{direct} qualifiers for ICF code $q$, i.e. they are directly linked to $q$.
	$TE_{m1}$ is the age of a qualifier in full days and $r_{m1}$ is the reliability of the linkage.
	\item \textbf{Direct input values for ICF code q:} These values ($m_{i}$, $\tilde{\alpha}_{mi}$, $\tilde{r}_{mi}$) are attached to the ICF code $q$. While $m_{i}$, the qualifier, can be directly attached to the ICF code, other values need to be calculated based on the available data.
	%\item \textbf{Child ICF codes with no measurements:} ICF codes from $ch.q(3)$ to $ch.q(k)$ are the child ICF codes of ICF code $q$ with no measurements linked to them. In this case they would be \textit{b28011}, \textit{b28012}, \textit{b28014}, \textit{b28015}, \textit{b28016}, \textit{b28018} and \textit{b28019}. Thus, in this case $k=9$.
	\item \textbf{Indirect qualifiers (I) for ICF code q:} This element depicts the two available qualifiers $s_{1}$ and $s_{2}$ for the two child ICF codes $ch1_{q}$ and $ch2_{q}$. $s_{1}$ and $s_{2}$ are the qualifiers for child ICF codes of ICF code $q$, i.e. they are the \textbf{indirect} qualifiers for ICF code $q$. The indirect qualifiers are linked to $q$ through its child ICF codes. $TE_{s1}$ is the time elapsed value and $r_{s1}$ is the reliability of the source linkage.
	\item \textbf{Indirect input values for ICF code q:} These values ($s_{j}$, $\tilde{\alpha}_{sj}$, $u_{sj}$, $\tilde{r}_{sj}$) are attached to the child ICF codes, and are used to calculate the values, $s_{ch1(q)}$ and $s_{ch2(q)}$, for the child ICF codes.  
\end{itemize}

There are three outputs in the diagram:

\begin{itemize}
	\item \textbf{Final calculated value for the ICF code q:} $x_{q}$ is the final value for ICF code $q$. It consists of \textbf{direct} qualifiers and \textbf{indirect} qualifiers made for the child ICF codes. $x_{q}$ is the weighted average of all available qualifiers: $s_{1}$, $s_{2}$, $m_{1}$ and $m_{2}$. In this case it would be calculated as follows: $x_{q}=(\tilde{\beta}_{m1}^{D}m_{1}+\tilde{\beta}_{m2}^{D}m_{2})+s_{ch1(q)}+s_{ch2(q)}=(\tilde{\beta}_{m1}^{D}m_{1}+\tilde{\beta}_{m2}^{D}m_{2})+(\tilde{\beta}_{s1}^{I}s_{1}+\tilde{\beta}_{s2}^{I}s_{2})+(\tilde{\beta}_{s2}^{I}s_{2})$, where $\beta$ terms refer to weighting coefficients. Details of the equation are explained in Section~\ref{subsec:calculation_of_ICF_code_and_r_values}.
	\item \textbf{Time weighting value for q:} $\alpha_{q}$ is the weighted mean of all available time weighting $\alpha$ values defined in Eq.~\ref{eq:weighted_mean_alpha_and_r}.
	\item \textbf{Linkage reliability value for q:} $r_{q}$ is the weighted mean of all available $r$ values, defined also in Eq.~\ref{eq:weighted_mean_alpha_and_r}.
\end{itemize}

\subsection{Calculation of ICF code and reliability values}
\label{subsec:calculation_of_ICF_code_and_r_values}

In the previous section, we explained computation of the ICF code value through an example for better understanding. In this section, we describe calculation of ICF code and realibility values more generally.

\subsubsection{ICF code value calculation}
\label{subsubsec:icf_code_value_calculation}

The value of the ICF code $q$ is defined as:

\begin{equation}\label{eq:qx}
	\begin{aligned}
		x_{q} &= f(\sum_{i \in D}^{}\tilde{\beta}_{mi}^{D}m_{i}+\sum_{k \in ch_{q}}^{} s_{chk(q)}+\sum_{k \in ch_{q}}^{} x_{chk(q)})\\
		\quad&=f(\sum_{i \in D}^{}\tilde{\beta}_{mi}^{D}m_{i}+\sum_{k \in ch_{q}}^{} \sum_{j \in k}^{}\tilde{\beta}_{sj}^{I}s_{j}+\sum_{k \in ch_{q}}^{} \tilde{\beta}_{xk}^{I}x_{k}).
	\end{aligned}
\end{equation}

This definition consists of four separate elements:

\begin{enumerate}
	\item Function $f$ is a weighting function that can be selected to either give emphasis to higher qualifiers (exponential weighting) in data or to highlight lower qualifiers (logarithmic weighting). It is also possible to use no weighting at all (linear weighting). The different weighting functions are presented in more detail in \ref{app:value_weighting_functions}.
	\item The first sum term  $\sum_{i \in D}^{}\tilde{\beta}_{mi}^{D}m_{i}$ is the weighted sum of qualifiers $m_{i}$ that are directly linked with the ICF code $q$. In the equation, $D$ refers to a set of \textbf{direct qualifiers}, i.e. those qualifiers that are directly linked to $q$. In other words, they are the direct qualifiers from the viewpoint of $q$.
	Additionally, we define a total weighting term $\beta^{D}$ for direct qualifiers:
	
	\begin{equation}\label{eq:beta_m}
		\beta_{mi}^{D}=\tilde{\alpha}_{mi}\tilde{r}_{mi}.
	\end{equation}
	
	As can be seen, the total weighting consists of two elements $\alpha$ and $r$. The first element $\alpha$ is a time weighting value that depends on age of the corresponding qualifier. Generally, as a qualifier gets older, the less weight it is given in the calculation. Time weighting is discussed in more detail in \ref{app:time_weighting}. The second element $r$ is the linkage reliability that we defined in Section~\ref{subsec:linkage_of_items}. It should be noted that $\alpha$, $r$ and $\beta$ in Eq.~\ref{eq:qx} carry the tilde notation ($\tilde{\alpha}$, $\tilde{r}$, $\tilde{\beta}$). This notation means that a normalized value for the term is used instead of the raw value. A standard normalization procedure is carried out to make the different values directly comparable. The two normalization equations are presented in \ref{app:normalization}.
	\item The second sum term $\sum_{k \in ch_{q}}^{} \sum_{j \in k}^{}\tilde{\beta}_{sj}^{I}s_{j}$ is the weighted sum of qualifiers $s_{j}$ that are indirectly, i.e. via the child ICF codes of $q$, linked with the ICF code $q$. In the outer sum, $ch_{q}$ refers to a child ICF code of $q$ that has a qualifier or a calculated value. Thus, each child ICF code of $q$ is utilized in the calculation. The inner sum runs through all the qualifiers linked to the child ICF code.
	A total weighting term for indirect qualifiers is called $\beta^{I}$. It is a slightly modified version from the weighting term shown in Eq.~\ref{eq:beta_m}, defined as:
	
	\begin{equation}\label{eq:beta_s}
		\beta_{sj}^{I}=\tilde{\alpha}_{sj}\tilde{r}_{sj}u_{sj},
	\end{equation}
	
	where $u_{sj}$ is the uniqueness of source of the qualifier $s_{j}$, calculated based on how many child ICF codes the same qualifier is linked to. It is defined as $u_{sj}=1/z_{j}, z_{j} \in \mathbb{N}^{+}$, where $z_{j}$ is the number of linkages between $s_{j}$ and child ICF codes of $q$. $u$ is designed to lower the qualifier weighting if the same source is used to form values for two or more 
	ICF codes under the same parent ICF code. $\alpha$ and $r$ have the same purpose as in Eq.~\ref{eq:beta_m}.
	\item The third sum term $\sum_{k \in ch_{q}}^{} \tilde{\beta}_{xk}^{I}x_{k}$ is similar to the second sum term above. The difference is that it runs through the calculated values ($x$), instead of the qualifiers ($s$). Here the second sum term is not needed, as each child ICF code can have only one calculated value. In addition, uniqueness of source is not applicable here. Hence, $\tilde{\beta}_{xk}^{I}$ is defined similarly to Eq.~\ref{eq:beta_m}, as $\tilde{\beta}_{xk}^{I}=\tilde{\alpha}_{xk}\tilde{r}_{xk}$.
\end{enumerate}

\subsubsection{ICF code reliability calculations}
\label{subsubsec:icf_code_reliability_calculations}

There are two values, $\alpha_{q}$ (equivalent of $\alpha_{level.index}$ on Alg.~\ref{alg:hi}, line \ref{alg1:line:calculate_alpha}, the weighted mean of time weighting $\alpha$) and $r_{q}$ (equivalent of $r_{level.index}$ on line \ref{alg1:line:calculate_r}, the weighted mean of linkage reliability $r$), that are calculated for $x_{q}$. They are defined almost identically to $x_{q}$ in Eq.~\ref{eq:qx}:

\begin{equation}\label{eq:weighted_mean_alpha_and_r}
	\alpha_{q}=\sum_{i \in D}^{}\tilde{\beta}_{mi}^{D} \alpha_{mi} + \sum_{k \in ch_{q}}^{} \sum_{j \in k}^{}\tilde{\beta}_{sj}^{I} \alpha_{sj} + \sum_{k \in ch_{q}}^{} \tilde{\beta}_{xk}^{I} \alpha_{xk},
\end{equation}

and similarly for $r_{q}$, by replacing the $\alpha$ term in the equation with $r$. Here the $\tilde{\beta}$ terms are used to determine the weightings for each $\alpha/r$ element. These values, $\alpha_{q}$ and $r_{q}$, ''travel'' with the $x_{q}$ value and are associated with it.

\section{Statistical analyses}
\label{sec:results}

The set of calculated health indices was validated by comparing the results with person's self-reported EQ-VAS answers and maximum pain level answers.
Two groups were formed for validation. For the first group the treatment period duration 
was at least 90 days, whereas for the second group the same limit was 30 days. Length of treatment sequence (see Section~\ref{subsec:overview_of_data}) in the first group was at least 10 days. For the 30-day group, the length of treatment sequence was at least five days. A person could be assigned to 
both groups, provided that the day and qualifier preconditions were satisfied.

To examine the effect of different time weightings, Pearson correlation coefficients were calculated for different time decay constants $\gamma$ (see \ref{app:time_weighting} for detailed definition of $\gamma$).
The values for $\gamma$ were chosen so that:

\begin{itemize}
	\item $\gamma_{1} = (1/20)^{1/30} \approx 0.905$ represented a very heavy time decay: a 30 day old qualifier was weighed only 5 \% of its original time weight.
	\item $\gamma_{2} = (1/3)^{1/30} \approx 0.964$ represented our estimated preliminarily potential time decay value in a real-life scenario. In practice this value means that a measurement made 30 days ago was weighed as one third of its original 
	time weight. 
	\item $\gamma_{3}=1$ means that there was no time decay at all.
\end{itemize}

The algorithm was implemented using Python programming language \cite{vanrossum2009python}. The statistical analyses were made using SciPy \cite{virtanen2020scipy} and figures with Matplotlib \cite{hunter2007matplotlib}.

\subsection{EQ-VAS answer vs. the health index}
\label{subsec:eq_vas_vs_hi}

First, the health index was calculated without giving any emphasis to lower or higher qualifiers. In Table~\ref{tbl:corr_info_linear_case} the EQ-VAS answer was compared with the health index, and Pearson correlations are presented for both of the groups. The influence of changing the $\gamma$ was also explored. The results show mostly moderate positive correlations for all configurations between the EQ-VAS answer and the calculated health index. For the 90-day group there were 84 persons, who had in total 125 EQ-VAS answers recorded. Most persons had only one EQ-VAS answer recorded, but some had two, three or even four answers. All of these answers were used in the calculations. The 30-day group had 115 persons with 159 EQ-VAS answers.

\begin{table}[h!]
	\centering
	\small
	\begin{tabular}{lrr}
		\hline
		{$\gamma$} &    30-day group &    90-day group\\
		\hline
		$(\frac{1}{20})^{1/30}$ & 0.690*** & 0.642*** \\
		$(\frac{1}{3})^{1/30}$  & 0.700*** & 0.659*** \\
		$1$                   & 0.654*** & 0.599*** \\
		\hline
	\end{tabular}
	\caption{Pearson correlations for the EQVAS answer vs. Health index. Significance of the correlation coefficients is indicated by stars: *** indicates $p<0.001$.}
	\label{tbl:corr_info_linear_case}
\end{table}

\subsection{Maximum pain vs. the health index}
\label{subsec:maximum_pain_vs_hi}

In the case of maximum pain versus the health index, the Pearson correlation was calculated individually for each person between the maximum pain trajectory and the health index trajectory.
The maximum pain is defined as the single highest value in the four pain level answers. A maximum pain trajectory was then formed for each person.
Since this information was also used in calculating the health index, 
there was inevitably at least some correlation between the two values. However, this comparison provides a useful second perspective to the validation of the model. 

Bonferroni correction is a method designed to prevent the data from incorrectly showing significance. The method is applied here in the maximum pain vs. health index trajectory correlation calculations, and $\alpha$ level is divided by $n$. 

\begin{figure}[h!]
	\centering
	\subfloat[90-day group.]{{\includegraphics[width=6.3cm]{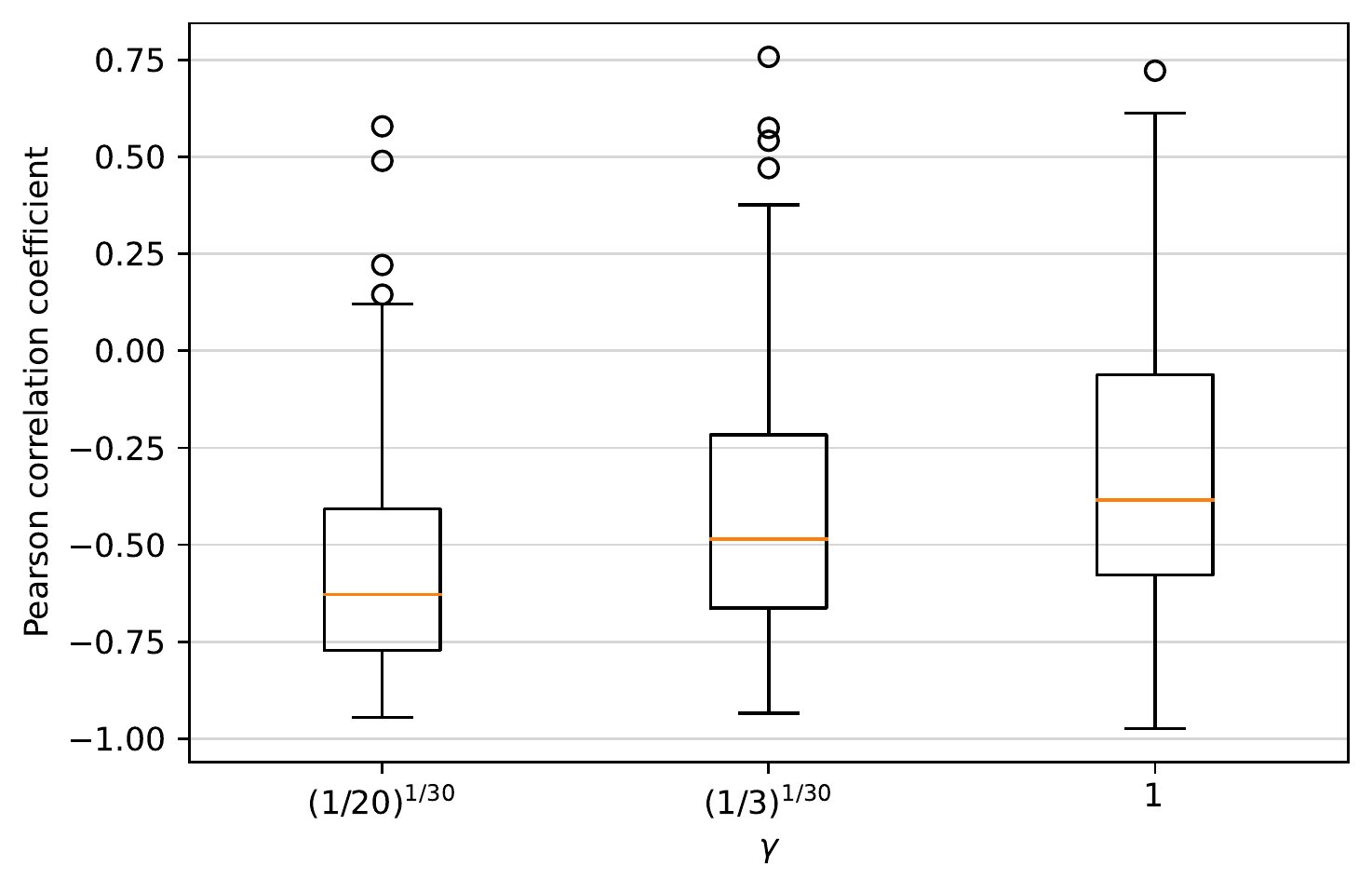} }\label{fig:hi_vs_maxpain_boxplots_90_10}}%
	\qquad
	\subfloat[30-day group.]{{\includegraphics[width=6.3cm]{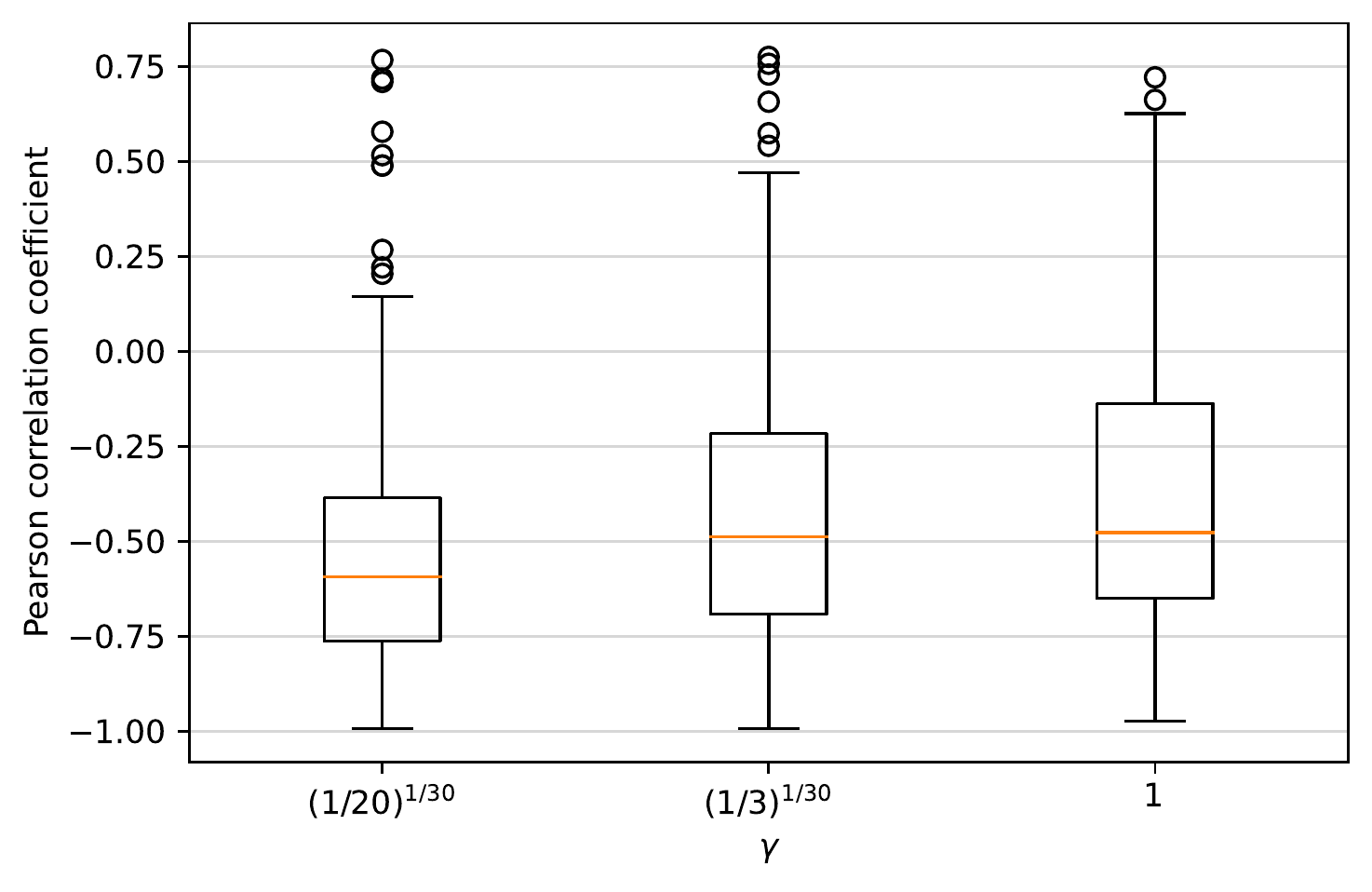} }\label{fig:hi_vs_maxpain_boxplots_30_5}}%
	\caption{Boxplots of Pearson correlations between the health index and maximum pain trajectories. \protect\subref{fig:hi_vs_maxpain_boxplots_90_10} shows the 90-day group, where correlation medians for the three time decay groups ($n=133$) are -0.628, -0.486, and -0.385, respectively. \protect\subref{fig:hi_vs_maxpain_boxplots_30_5} shows the 30-day group, where correlation medians for the three time decay groups are -0.594, -0.489 ($n=184$), and -0.477 ($n=182$).}%
	\label{fig:hi_vs_painvas_boxplot}%
\end{figure}

Distribution of correlation values for the two groups are presented in Fig.~\ref{fig:hi_vs_painvas_boxplot}. A slight upward trend in median correlation is visible in both groups: as $\gamma$ increases, the median correlation gets closer to zero. In all cases, the correlation was negative, varying roughly from low to moderate. There are small differences in the $n$ values for the different $\gamma$ groups. In some configurations, the health index value remained constant for some persons, and thus calculating the correlation was not possible. These persons were omitted from the calculations.

Looking at the portions of significant correlations, for the 90-day group, Bonferroni corrected significant (*) portions were 55 \% for $\gamma_{1}$, 41 \% for $\gamma_{2}$ and 31 \% for $\gamma_{3}$. In the 30-day group, the similar portions were 41 \% for $\gamma_{1}$, 29 \% for $\gamma_{2}$ and 23 \% for $\gamma_{3}$.
Thus, there are more significant correlations when length of the treatment sequence and duration of the treatment period increase.

In Table~\ref{tbl:bin_stats} the effect of time series length is further examined by binning the maximum pain vs. health index trajectories of the persons into three bins based on the length of treatment sequence. These results also indicate that as more data points are included in the trajectory, significant portion increases.

\begin{table}[h!]
	\small
	\centering
	\begin{tabular}{lrlll} 
		\hline
		Bin & Day ranges             & \begin{tabular}[c]{@{}l@{}}Bonferroni signif-\\icant (*) portions\end{tabular} & \begin{tabular}[c]{@{}l@{}}Median \\correlations\end{tabular} & \begin{tabular}[c]{@{}l@{}}Subset \\lengths\end{tabular}  \\ 
		\hline
		1   & $[10,26]$, $[5,15]$    & 18.8 \%, 7.9 \%                                                                & -0.385, -0.569                                                & 48, 63                                                    \\
		2   & $[27,42]$, $[16,32]$   & 36.6 \%, 18.3 \%                                                               & -0.500, -0.376                                                & 41, 60                                                    \\
		3   & $[43,325]$, $[33,325]$ & 70.5 \%, 65.6 \%                                                               & -0.510, -0.511                                                & 44, 61                                                    \\
		\hline
	\end{tabular}
	\caption{Maximum pain vs. health index correlation statistics when using three bins with the length of treatment sequence as the binning criterion. The three bins are roughly the same size (tertiles). Here the time decay value was defined as $\gamma_{2}$. Both, the 90-day group and the 30-day group, are presented, separated by a comma. \textit{Day ranges} shows, for each bin, the lengths of treatment sequences included. For example, the \textit{Bin 1} includes persons with length of treatment sequence from 10 to 26 days (the 90-day group) or persons with length of treatment sequence from 5 to 15 days (the 30-day group). \textit{Bonferroni significant (*) portions} refers to the portion of Bonferroni corrected significant correlations at level 0.05. \textit{Median correlations} shows the median of all correlations in the bin, while \textit{Subset lengths} gives the number of persons used for each bin.}
	\label{tbl:bin_stats}
\end{table}

\begin{figure}[ht!]
	\centering
	\subfloat[Moderate negative correlation.]{{\includegraphics[width=6.3cm]{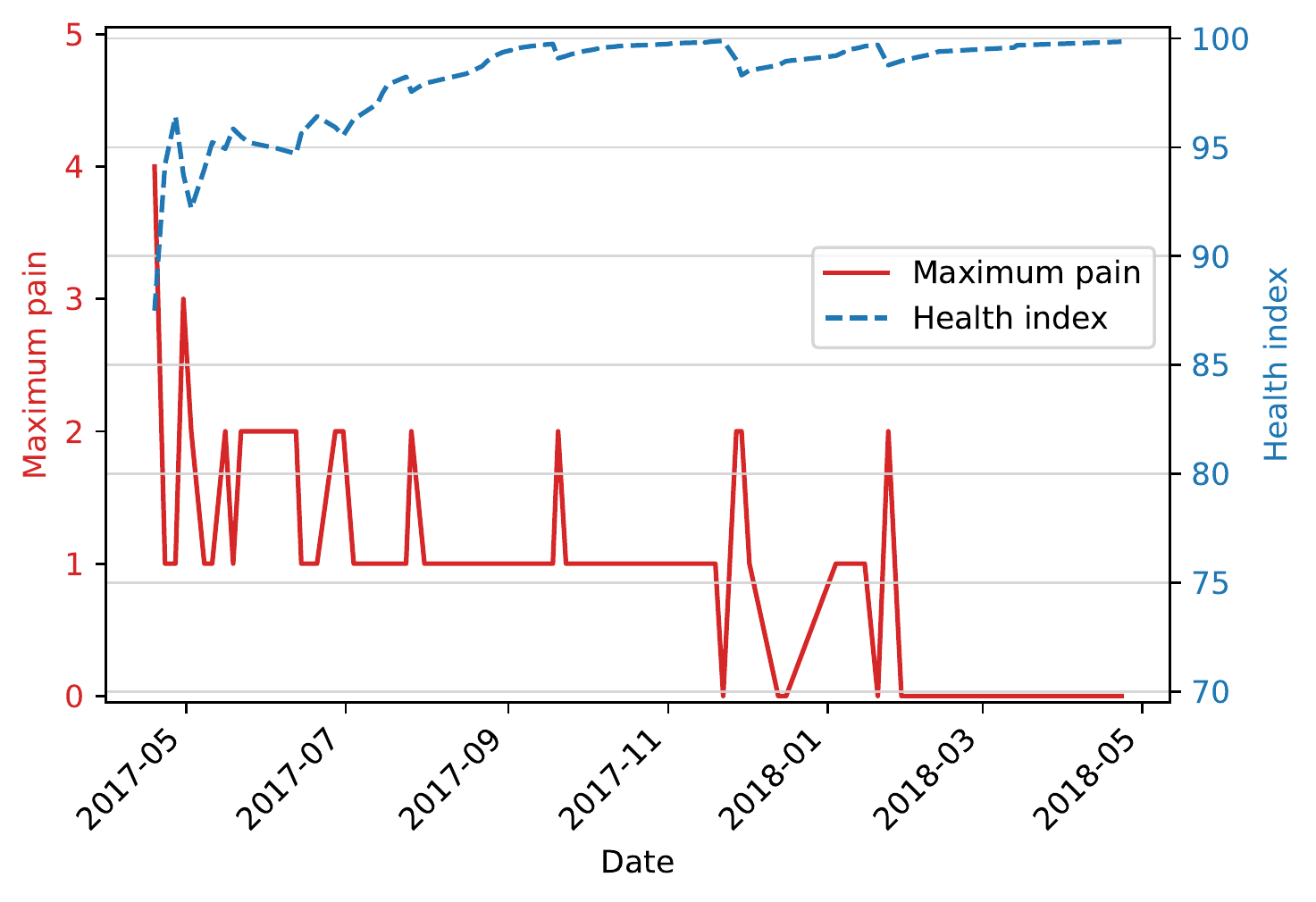} }\label{fig:hi_vs_maxpain_typical_case}}%
	\qquad
	\subfloat[Very high negative correlation.]{{\includegraphics[width=6.3cm]{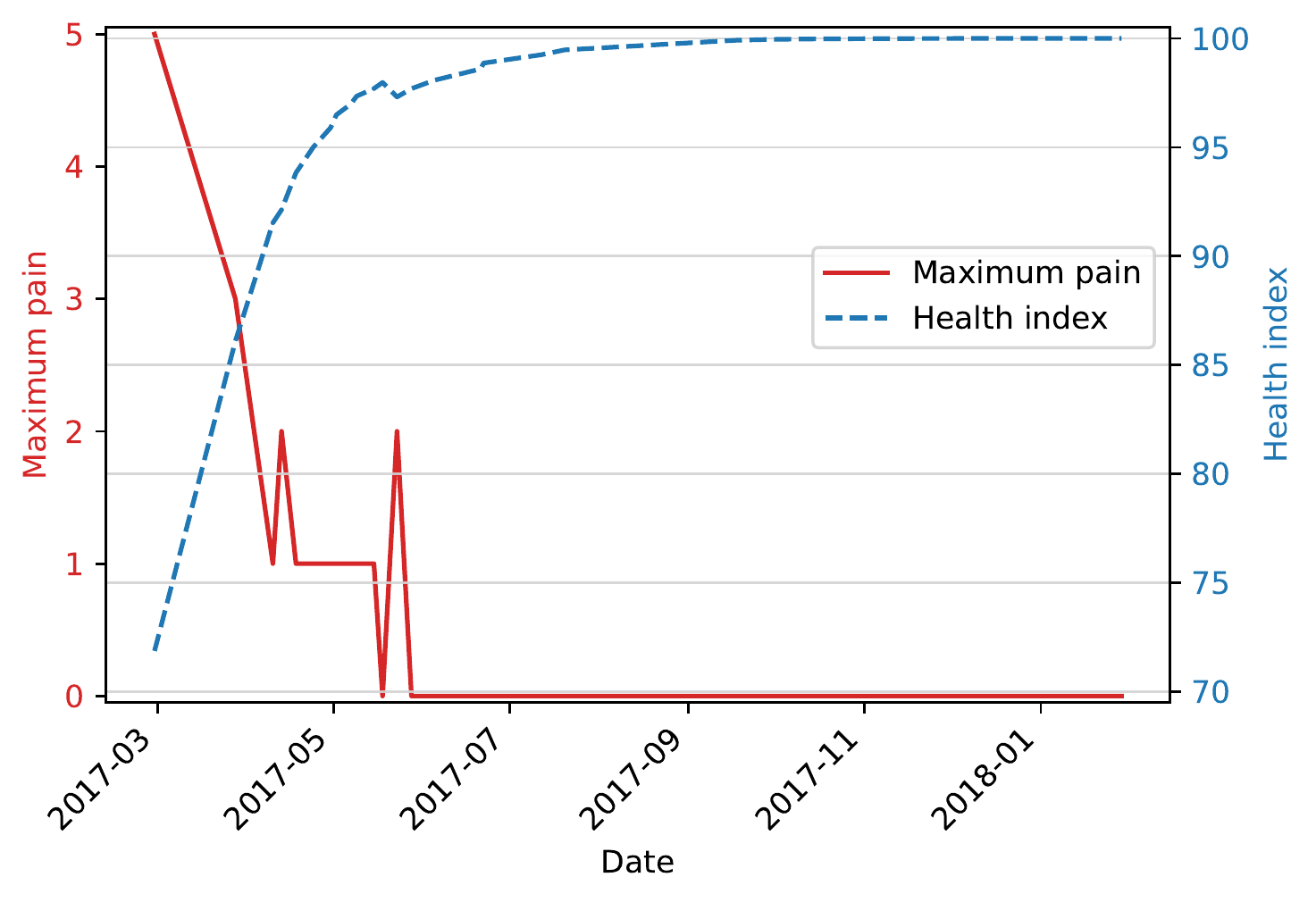} }\label{fig:hi_vs_maxpain_extreme_case}}%
	\caption{Two cases of correlations between the maximum pain and health index trajectories. \protect\subref{fig:hi_vs_maxpain_typical_case} is a typical case showing moderate negative correlation (-0.670) with a length of treatment sequence 40 days and \protect\subref{fig:hi_vs_maxpain_extreme_case} is an extreme case showing very high negative correlation (-0.934) with a length of treatment sequence 28 days. Time decay was defined as  $\gamma_{2}$ in both cases. Bonferroni corrected correlations were significant (***).}%
	\label{fig:hi_vs_maxpain_example_trajectories}%
\end{figure}

In Fig.~\ref{fig:hi_vs_maxpain_example_trajectories} two example persons and their maximum pain vs. health index trajectories are presented. Firstly, there is a typical example with moderate negative correlation between the trajectories. Secondly, trajectories showing very high negative correlation are presented.

\subsection{Effect of non-linear weighting}
\label{subsec:effect_of_non-linear_weighting}

In this section, the effect of qualifier weighting for correlations is examined. Qualifier weighting can be calibrated in the model by tuning the $y$ parameter. There are three different functions that the weighting can use: 1) when $y \in (0,2)$, the weighting is exponential, meaning that higher qualifier values have more weight, 2) when $y \in (2,4)$, the weighting is logarithmic; lower qualifier values have more weight, and 3) when $y=2$, the weighting is linear; qualifier values are treated without any weighting. Details on the value weighting can be found in \ref{app:value_weighting_functions}.

\begin{figure}[ht!]
	\centering
	\subfloat[90-day group ($n=125$).]{{\includegraphics[width=6.3cm]{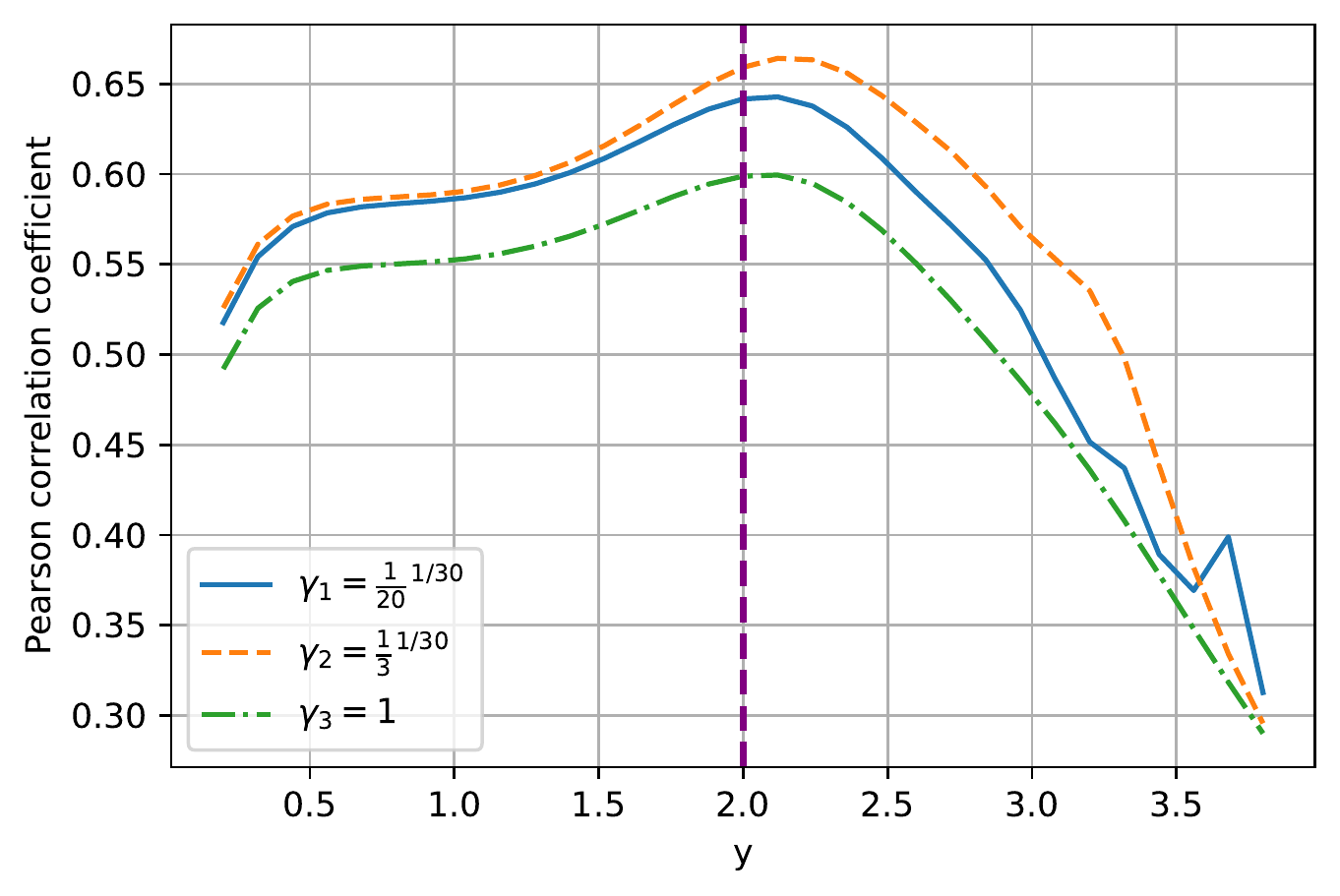} }}%
	\qquad
	\subfloat[30-day group ($n=159$).]{{\includegraphics[width=6.3cm]{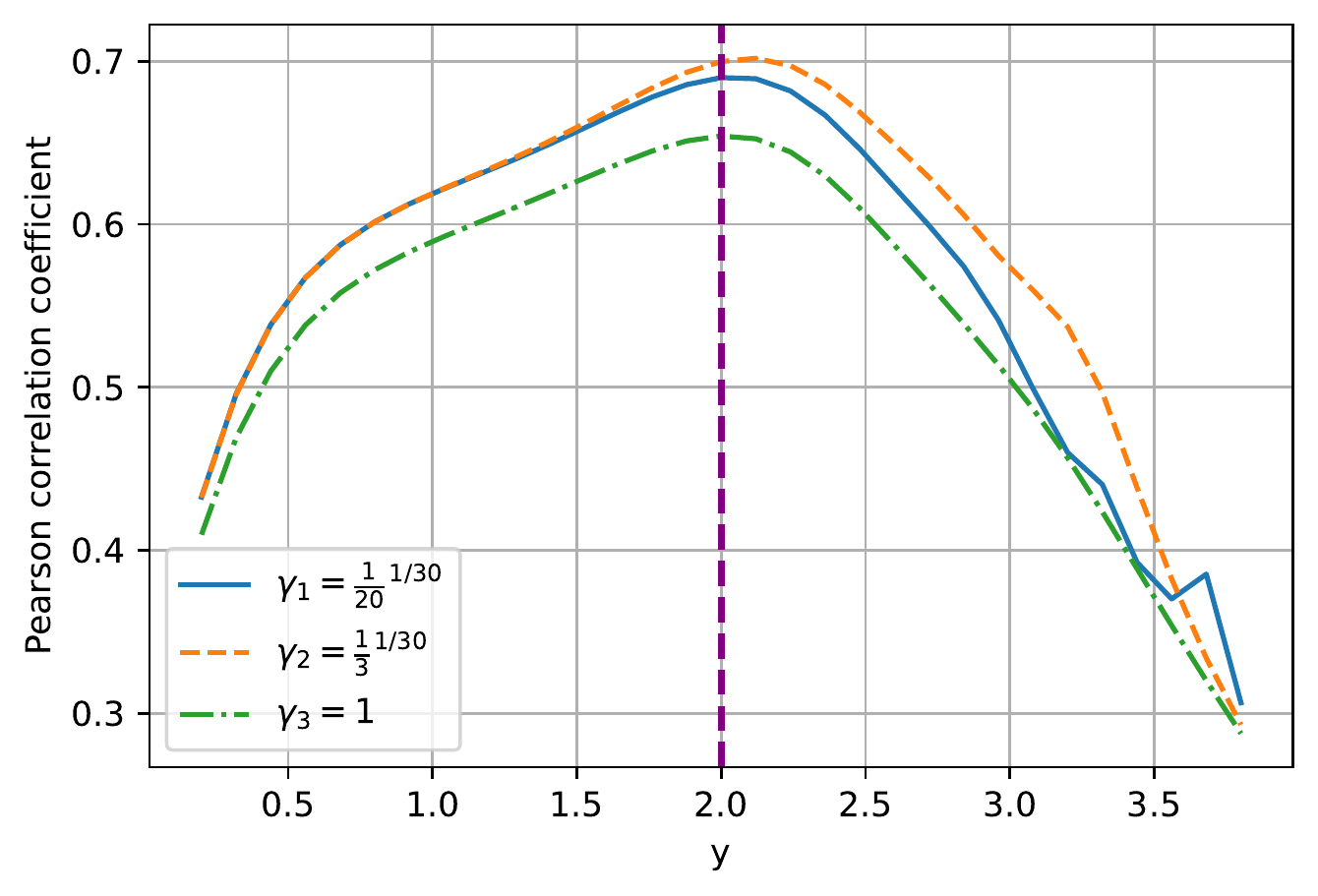} }}%
	\caption{Pearson correlation between the health index and EQ-VAS answer.}%
	\label{fig:hi_vs_eqvas_pearson}%
\end{figure}

For the $y$ parameter, 15 evenly spaced values were chosen for both sides of the linear ($y=2$) case, meaning in total $15+1+15=31$ $y$ parameter values were examined from $y=0.2$ to $y=3.8$ with intervals of $0.2$. 
The results for health index versus EQ-VAS answer using Pearson correlation are shown 
in Fig.~\ref{fig:hi_vs_eqvas_pearson}. For the 90-day group, highest correlations for all three $\gamma$ values were observed at $y=2.12$. The correlations were 0.643, 0.664, and 0.599, respectively. In the 30-day group the highest correlations were observed at the linear weighting point ($y=2$). Correlations at that point are the same presented in Table~\ref{tbl:corr_info_linear_case}.

\begin{figure}[ht!]
	\centering
	\subfloat[90-day group.]{{\includegraphics[width=6.3cm]{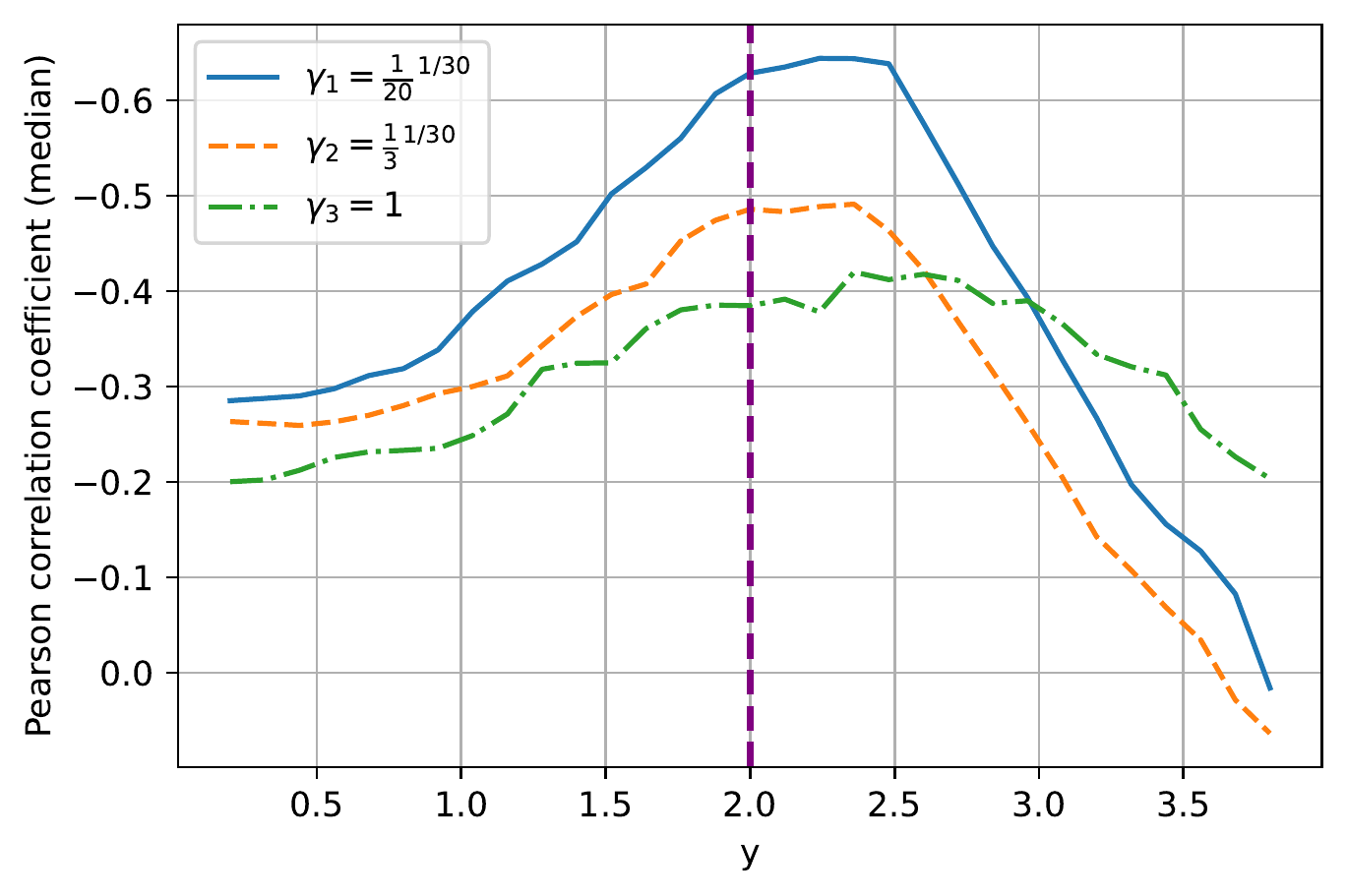} }\label{fig:hi_vs_max_pain_90_10_Y_GAMMA}}%
	\qquad
	\subfloat[30-day group.]{{\includegraphics[width=6.3cm]{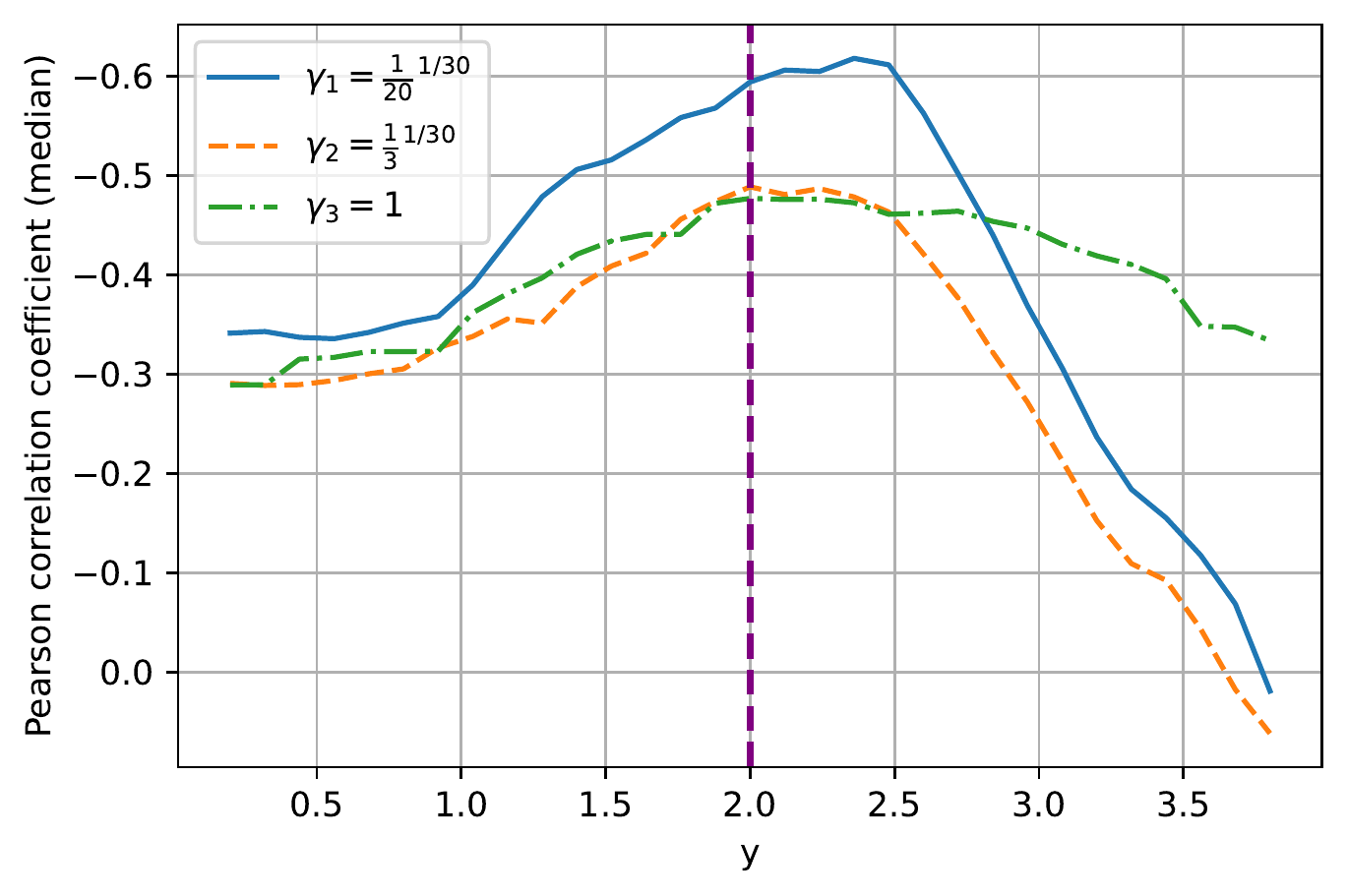} }\label{fig:hi_vs_max_pain_30_5_Y_GAMMA}}%
	\caption{Median Pearson correlation between the health index and maximum pain. \protect\subref{fig:hi_vs_max_pain_90_10_Y_GAMMA} shows the 90-day group ($\gamma_{1}, \gamma_{2}$:  $n=133$, $\gamma_{3}$:  $n=132,133$ ) and \protect\subref{fig:hi_vs_max_pain_30_5_Y_GAMMA} the 30-day group ($\gamma_{1}, \gamma_{2}$:  $n=184$, $\gamma_{3}$:  $n=181, 182, 183$ ).} %
	\label{fig:hi_vs_painvas_median}%
\end{figure}

In case of maximum pain vs. the health index, median correlations for the two groups are presented in Fig.~\ref{fig:hi_vs_painvas_median}. Here, the highest negative correlations observed for the three $\gamma$ values for the 90-day group were -0.644 (when $y=2.24$), -0.491 and -0.420 (both when $y=2.36$), respectively. In the 30-day group the correlations were -0.618 (when $y=2.36$), -0.489 and -0.477 (both when $y=2$).

\section{Discussion}
\label{sec:discussion}

\subsection{Conclusions}
\label{sec:conclusions}

The ICF framework provides an ideal platform for developing a health index, as it is a widely-used, standardized classification system that covers comprehensively aspects affecting person's health and functioning. Its validity has been proven by numerous published studies and it is heavily utilized worldwide for population health research projects \cite{madden:_the_icf_has_made_a_difference_to_functioning_and_disability_measurement_and_statistics}. Another advantage of using the ICF is that many questionnaires used to evaluate treatment response have been linked to ICF codes, allowing questionnaire results to be converted to ICF code qualifiers. These linkages have been scientifically validated and there are extensive guides \cite{world2013use, cieza2019} available for creating new linkages for newer datasets.

The proposed method can also be easily applied to calculate a 
health profile. In the context of the ICF, a health profile would most likely be a set of separate scores for each of the four ICF \textit{components} (a. Body functions and structures, b. Activities and participation, c. Environmental factors, d. Personal factors). It is also possible to create a more detailed profile by going deeper into the ICF structure.

This study established a framework for homogenization of data sets across clinics so that they can be
concatenated into a single very large data set. This step is highly crucial for development of AI,
because training of machine learning algorithms requires that same variables are available from all analyzed persons. If
homogenization of data sets is not performed, only subsets of persons and/or variables can be analyzed at once, reducing 
predictive power of the trained models.
When going a few steps further, one of the goals is to employ the health index as an optimization target in a machine learning model. 
This approach will make it possible to improve the person's health comprehensively instead of focusing on a single health parameter and can, for example, help in choosing the optimal rehabilitation pathway for a person.

In addition to being able to handle heterogenous data from various international sources, it is expected that the proposed index can be formed even when there are only few qualifiers available for a person. 
However, when there are more qualifiers available, the health index becomes more reliable.
One of the aims was also to emphasize recent qualifiers in health index calculation, while also giving some weight to qualifiers from the past.

\subsection{Limitations}
\label{subsec:limitations}

The ICF framework enables its users to define health and functioning of an individual in very detailed ways. In \textit{Body structures (s)} construct the second qualifier can be used to indicate the nature of the change in the respective body structure, and the third qualifier indicates the location, e.g. left or right \cite{who:_icf,world2013use}. Furthermore, in \textit{Activities and participation (d)} the second generic qualifier (capacity) is used to indicate limitation without assistance \cite{who:_icf}. The proposed model utilizes only the first generic qualifier. Thus, any further information potentially provided by these additional qualifiers is not utilized in the current model.

In the case of ICF's \textit{Environmental factors (e)} construct the first qualifier can be used to denote either the positive effects (facilitators) or the negative effects (barriers) of the environment \cite{who:_icf}. The model is currently able to handle only barriers. Thus, the potential positive effect of possible facilitators to the health index is not part of the proposed model.

Qualifiers 8 (not specified) and 9 (not applicable) are not handled by the proposed model. Currently the values are treated as ordinal, and these nominal values can not be utilized. This limitation might result in loss of some information, since these qualifiers are used in ICF system to record information that can not be captured using the $[0,4]$ value range. Qualifier 8 should be applied when there is a problem, but it is unknown whether that problem is mild or severe. Additionally, qualifier 9 is typically used in a situation when use of the category is not appropriate for the individual \cite{world2013use}.

A potential limitation regarding the practical usability of the model is that too scarce data might give a wrong impression about person's overall health. The proposed model does not currently specify any metrics for observing how complete the employed data is. Any personal factors that are not expressed by the qualifier format, such as age of a person, can not be utilized in the current calculation process.

\section*{CRediT authorship contribution statement}

\textbf{Ilkka Rautiainen:} Methodology, Software, Validation, Formal analysis, Investigation, Writing - Original Draft, Visualization, \textbf{Lauri Parviainen:} Conceptualization, Methodology, Data Curation, \textbf{Veera Jakoaho:} Data Curation, Writing - Review \& Editing, Project administration, \textbf{Sami \"{A}yr\"{a}m\"{o}:} Conceptualization, Methodology, Writing - Review \& Editing, Supervision, \textbf{Jukka-Pekka Kauppi:} Conceptualization, Methodology, Software, Writing - Review \& Editing, Visualization, Supervision

\section*{Declaration of Competing Interest}

The authors declare that they have no known competing financial interests or personal relationships that could have appeared to influence the work reported in this paper.

\section*{Acknowledgements}

Ilkka Rautiainen's work was funded by David Health Solutions Ltd. and Jenny and Antti Wihuri Foundation (grant numbers 00180312, 00200302, and 00210295).

\bibliographystyle{elsarticle-num}
\bibliography{utilizing_the_icf_in_forming_a_personal_health_index_sources}

\begin{appendices}

\section{Details on data linkages}
\label{app:data_linkage_tables}
% format the table environment
\setcounter{table}{0}
\renewcommand{\thetable}{A.\arabic{table}}

The Oswestry low back pain disability questionnaire (ODI) includes one item on pain and nine items on activities of daily living (lifting, walking, social life, personal care, sitting, standing, sleeping, traveling, and sex life), each scored on a 0--5 scale, 5 representing the highest disability \cite{kocc2018comparison}. We defined the linkages ourselves, and they are depicted in Table~\ref{tbl:odi_linkages}. Further, the translations between the original ODI answer and the equivalent value as an ICF code qualifier are shown in Table~\ref{tbl:odi_vals}. The original ODI answer refers to the selected response number.

\begin{table}[ht!]
	\small
	\centering
	\begin{tabular}{lll}
		\hline
		\textbf{Item as appeared}                                                        & \begin{tabular}[c]{@{}l@{}}\textbf{Purpose of information }\\\textbf{in question}\end{tabular} & \textbf{ICF code(s)}  \\
		\hline
		Pain intensity                                                                   & Level of pain                                                                                  & b280                  \\
		\begin{tabular}[c]{@{}l@{}}Personal care \\(washing, dressing etc.)\end{tabular} & \begin{tabular}[c]{@{}l@{}}Pain related to personal care \\tasks\end{tabular}                  & b280, d5              \\
		Lifting                                                                          & Pain related to lifting objects                                                                & b280, d430            \\
		Walking                                                                          & Pain related to walking                                                                        & b280, d450            \\
		Sitting                                                                          & Pain related to sitting                                                                        & b280, d4103           \\
		Standing                                                                         & Pain related to standing                                                                       & b280, d4104           \\
		Sleeping                                                                         & Pain related to sleep                                                                          & b280, b1340           \\
		Sex life (if applicable)                                                         & Pain related to sexual activities                                                              & b280, d7702, b640     \\
		Social life                                                                      & \begin{tabular}[c]{@{}l@{}}Pain related to participation \\in social activities\end{tabular}   & b280, d910            \\
		Travelling                                                                       & \begin{tabular}[c]{@{}l@{}}Pain related to capacity for \\travelling\end{tabular}              & b280, d470           \\
		\hline
	\end{tabular}
	\caption{The ODI questions, their purpose, and their equivalent ICF codes. The response options are not listed.}\label{tbl:odi_linkages}
\end{table}

\begin{table}[ht!]
	\small
	\centering
	\begin{tabular}{l|c|c|c|cc|c|}
		\cline{2-7}
		Original ODI answer: & 0 & 1 & 2 & \multicolumn{1}{c|}{3} & 4 & 5 \\ \cline{2-7} 
		ICF code qualifier:  & 0 & 1 & 2 & \multicolumn{2}{c|}{3}     & 4 \\ \cline{2-7} 
	\end{tabular}
	\caption{The original ODI answers and their translated equivalent values in the ICF system.}\label{tbl:odi_vals}
\end{table}

The five-level version of the EQ-5D generic health questionnaire, EQ-5D-5L consists of five questions on mobility, self-care, usual activities, pain/discomfort, and anxiety/depression. The scale of answers is on a 1--5 scale, with 5 being the highest disability. In addition to the five questions, there is a self-assessed score for overall health on a 0--100 scale, EQ-VAS. It is not used when calculating the health index. Instead, EQ-VAS is only used for the external validation of the index. Since, similarly to the 0--4 scale of the ICF qualifier, there are five response options in all the questions excluding the EQ-VAS. Thus, the answers can be directly mapped as ICF code qualifiers. The questions and their linkages to ICF codes are described in Table~\ref{tbl:eq5d_linkages}. We defined these linkages ourselves.

\begin{table}[ht!]
	\small
	\centering
	\begin{tabular}{lll}
		\hline
		\textbf{Item as appeared}                                                                                                & \begin{tabular}[c]{@{}l@{}}\textbf{Purpose of information }\\\textbf{in question}\end{tabular}         & \textbf{ICF code(s)}  \\
		\hline
		Mobility                                                                                                                 & Problems with walking                                                                                  & d450, d455            \\
		Self-care                                                                                                                & \begin{tabular}[c]{@{}l@{}}Problems with self-care, \\specifically washing and \\dressing\end{tabular} & d5, d510, d540        \\
		\begin{tabular}[c]{@{}l@{}}Usual activities (e.g. work, \\study, housework, family \\or leisure activities)\end{tabular} & \begin{tabular}[c]{@{}l@{}}Problems with daily routine \\/ usual tasks\end{tabular}                    & d230                  \\
		Pain / Discomfort                                                                                                        & Level of pain or discomfort                                                                            & b280                  \\
		Anxiety / Depression                                                                                                     & \begin{tabular}[c]{@{}l@{}}Level of anxiety or \\depression\end{tabular}                               & b152, b1528           \\
		\begin{tabular}[c]{@{}l@{}}We would like to know how \\good or bad your health is \\TODAY.\end{tabular}                  & General health perception                                                                              & N/A                  \\
		\hline
	\end{tabular}
	\caption{The EQ-5D-5L questions, their purpose, and their equivalent ICF codes. The response options are not listed.}\label{tbl:eq5d_linkages}
\end{table}

Pain levels in the back, hip/leg, neck, and shoulder/arm areas were measured from persons in the visual analogue scale (VAS) on a 0--10 scale in the beginning of every visit to a clinic. Because these variables were measured frequently, it is important to include these data sets in the analysis besides questionnaire data, which was collected typically only once or twice during a whole treatment period.
We mapped all the four pain level answers ourselves into ICF.

In Table~\ref{tbl:pain_level_linkages} the linkages between the original different pain level questions and the ICF codes are shown. Further, the translations between the original pain level answer and the equivalent value as an ICF code qualifier are shown in Table~\ref{tbl:pain_level_vals}. For example, the original pain level answer of ''3'' for back pain would translate to ICF code $b28013$ as qualifier ''1''.

\begin{table}[ht!]
	\small
	\centering
	\begin{tabular}{ll}
		\hline
		\textbf{Pain level} & \textbf{Corresponding ICF code} \\
		\hline
		Back                & b28013 (Pain in back)           \\
		Hip/leg                 & b28015 (Pain in lower limb)     \\
		Neck                & b28010 (Pain in head and neck)  \\
		Shoulder/arm        & b28014 (Pain in upper limb)  \\ \hline
	\end{tabular}
	\caption{The original pain level questions and their equivalent ICF codes.}\label{tbl:pain_level_linkages}
\end{table}

\begin{table}[ht!]
	\small
	\centering
	\begin{tabular}{l|cc|cc|ccc|cc|cl|}
		\cline{2-12}
		Original pain level: & \multicolumn{1}{c|}{0} & 1 & \multicolumn{1}{c|}{2} & 3 & \multicolumn{1}{c|}{4} & \multicolumn{1}{c|}{5} & 6 & \multicolumn{1}{c|}{7} & 8 & \multicolumn{1}{c|}{9} & 10 \\ \cline{2-12} 
		ICF code qualifier:  & \multicolumn{2}{c|}{0}     & \multicolumn{2}{c|}{1}     & \multicolumn{3}{c|}{2}                              & \multicolumn{2}{c|}{3}     & \multicolumn{2}{c|}{4}      \\ \cline{2-12} 
	\end{tabular}
	\caption{The original pain level answers and their translated equivalent values in the ICF system.}\label{tbl:pain_level_vals}
\end{table}

We also created new linkages for the mobility/strength tests. Strength and mobility levels related to relevant spine functions were measured using David Health Solutions' spine concept rehabilitation machines and presented in newton-metres and degrees, respectively. These values were converted to relative changes with respect to the average values of a reference population. Better mobility/strength than reference was mapped to 0 \%, which corresponded to \textit{no problem}. The interventions employed with the machines as well as their ICF codes are described in Table~\ref{tbl:control_test_linkages}. The translations of the original relative change values to the ICF code qualifiers are shown in Table~\ref{tbl:control_test_vals}. The values were then converted to the ICF domain employing the linkages depicted using the information in these two tables. For example, the relative change of 20 \% observed through the f120 machine would be linked with qualifier ''1'' to the ICF codes $b780$, $b7305$, $b7355$ and $b7401$.

\begin{table}[ht!]
	\scriptsize
	\centering
	\begin{tblr}{
			cell{2}{3} = {r=2}{},
			cell{4}{3} = {r=2}{},
			cell{6}{3} = {r=2}{},
			vlines,
			hline{1-2,4,6,8} = {-}{},
			hline{3,5,7} = {1-2}{},
		}
		\textbf{Intervention}                            & \textbf{Target of intervention}                                                                   & \textbf{Corresponding ICF codes}                                                                                                                                                                                                                                         \\
		110 Trunk Extension                              & {Increase of dorsal, \\lumbar and thoracic \\region muscle tone \\and strength, \\Pain prevention   } & {b7305 (Power of muscles of the trunk)\\b7355 (Tone of muscles of trunk)\\b7401 (Endurance of muscle groups)\\b780 (Sensations related to muscles \\and movement functions)}                                                                                             \\
		130 Trunk Flexion                                & {Increase of abdominal \\region muscle \\tone and strength, \\Pain prevention   }                   &                                                                                                                                                                                                                                                                          \\
		120 Trunk Rotation                               & {Increase of lateral and \\abdominal region \\muscle tone and strength,\\Pain prevention    }       & {b7302 (Power of muscles of one side \\of the body)\\b7305 (Power of muscles of the trunk)\\b7355 (Tone of muscles of trunk)\\b7401 (Endurance of muscle groups)\\b780 (Sensations related to muscles \\and movement functions)          }                               \\
		{150 Trunk Lateral \\Flexion    }                & {Increase of lateral and \\abdominal region \\muscle tone and strength,\\Pain prevention    }       &                                                                                                                                                                                                                                                                          \\
		{140 Cervical Extension / \\Lateral Flexion    } & {Increase of cervical region \\muscle tone \\and strength, \\Pain prevention    }                   & {b7300 (Power of isolated muscles \\and muscle groups)\\b7302 (Power of muscles of one side \\of the body)\\b7350 (Tone of isolated muscles and \\muscle groups)\\b7400 
			(Endurance of isolated muscles)\\b780 (Sensations related to muscles \\and movement functions)} \\
		160 Cervical Rotation                            & {Increase of cervical region \\muscle tone \\and strength, \\Pain prevention}                       &                                                                                                                                                                                                                                                                          
	\end{tblr}
	\caption{The original spine concept rehabilitation machine interventions, their purpose and their equivalent ICF codes.}\label{tbl:control_test_linkages}
\end{table}

\begin{table}[ht!]
	\scriptsize
	\centering
	\begin{tabular}{l|ll|lll|lll|llll|ll|}
		\cline{2-15}
		Original relative change (\%): & \multicolumn{2}{l|}{$0 \leq x \leq 4$} & \multicolumn{3}{l|}{$4 < x \leq 24$} & \multicolumn{3}{l|}{$24 < x \leq 49$} & \multicolumn{4}{l|}{$49 < x \leq 95$} & \multicolumn{2}{l|}{$95 < x \leq 100$} \\ \cline{2-15} 
		ICF code qualifier:            & \multicolumn{2}{c|}{0}                 & \multicolumn{3}{c|}{1}               & \multicolumn{3}{c|}{2}                & \multicolumn{4}{c|}{3}                & \multicolumn{2}{c|}{4}                 \\ \cline{2-15} 
	\end{tabular}
	\caption{The original spine concept rehabilitation machine relative changes (\%) and their translated equivalent values in the ICF system.}\label{tbl:control_test_vals}
\end{table}

\section{Defining weightings and normalization}
\label{app:defining_weightings}
%%% For Appendix B.
% format the equation environment
\renewcommand{\theequation}{B.\arabic{equation}}
% reset the counter
\setcounter{equation}{0}
% format the figure environment
\setcounter{figure}{0}
\renewcommand{\thefigure}{B.\arabic{figure}}

There are two separate weighting elements in the health index: time decay weighting and different value weighting functions. The purpose of the time weighting is to give more emphasis to newer qualifiers. By using the value weighting functions there is a possibility to tune measured values, either to give more emphasis to higher handicap levels or to downplay their role in the data.

\subsection{Value weighting functions}
\label{app:value_weighting_functions}

The weighting function is selected by first defining a tuning parameter $y\in]0,4[$ that determines the steepness of the weighting function. The curve always starts from point $(0,0)$, is fitted to go through point $(2,y)$ and always ends in point $(4,4)$. The selected $y$ value directly effects the function type used in fitting. There are three types of functions.

The function is exponential when $y \in (0,2)$, defined as:

\begin{equation}\label{eq:exp_weight}
	f(x)=ae^{bx}+c.
\end{equation}

The function is logarithmic when $y \in (2,4)$, defined as:

\begin{equation}\label{eq:log_weight}
	f(x)=a\textup{ ln}(bx+1).
\end{equation}

Finally, the function is linear when $y=2$. Linear function is simply defined as $f(x)=x$.

\begin{figure}[h!]
	\centering
	\includegraphics[height=4.6cm,keepaspectratio]{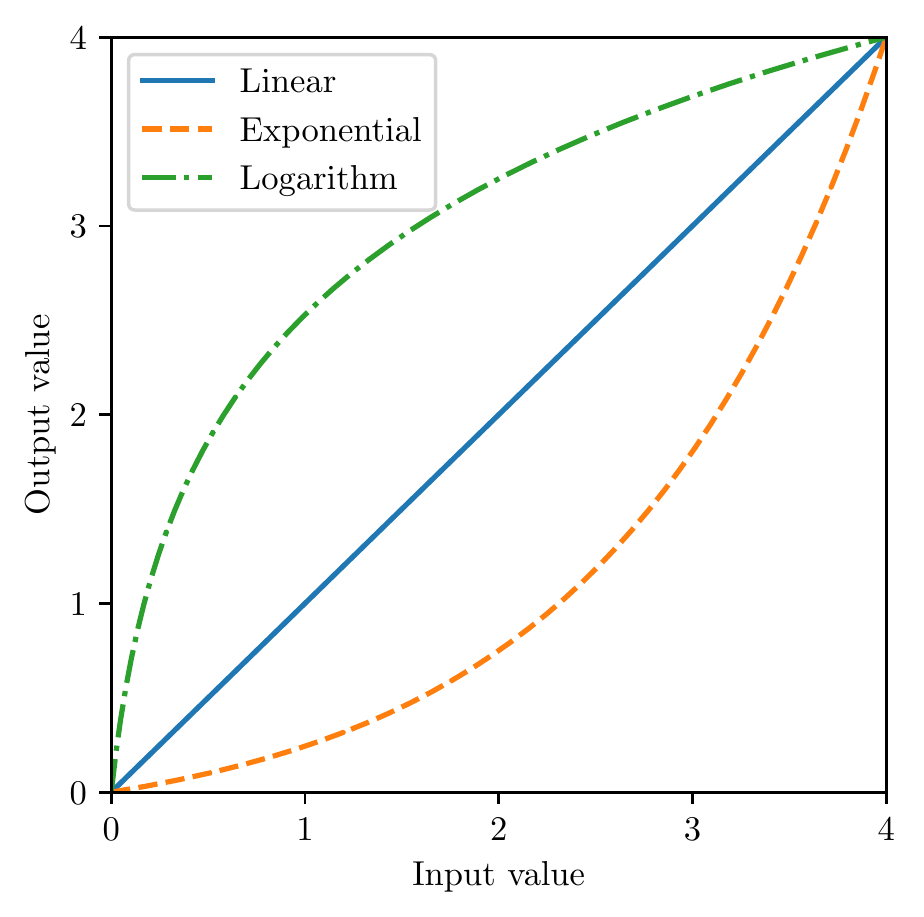}
	\caption{Weighting functions visualized using example $y$ values.}
	\label{fig:weighting_functions}
\end{figure}

The values for $a$, $b$ and $c$ in exponential and logarithmic functions are solved during the standard curve fitting process. For all weighting functions apply $x \in [0,4]$, meaning the range is the same as in the generic qualifier in the ICF. The three weighting functions are visualized in Fig.~\ref{fig:weighting_functions}, using values $y=0.75$ for the exponential and $y=3.25$ for the logarithm function. Furthermore, selecting exponential weighting function will mean that higher values, i.e. higher handicap levels, in data are emphasized. 
By contrast, when logarithmic weighting is employed, all input values are increased by the function, meaning that lower handicap levels are emphasized.

\subsection{Time weighting}
\label{app:time_weighting}

After linking the data to the ICF, the next step is to form a table that for each individual person contains all ICF code qualifiers available. 
In order to prepare the time series of different persons more easily comparable, the date of the person's any first qualifier is always defined as 0. 
Each consecutive measurement day for the person is then given a number $d \in \mathbb{N}$, expressing in full calendar days the distance from the start of the treatment. We can then define $TE$, the time elapsed in days from the latest valid qualifier:

\begin{equation}\label{eq:TE}
	TE=d-d_{0},
\end{equation}

where $d_{0}$ is the date of the person's latest valid qualifier, excluding $d$. Raw time weighting $\alpha$ is then defined as:

\begin{equation}\label{eq:alpha}
	\alpha=\gamma^{TE},
\end{equation}

where $\gamma \in (0,1]$ is the time decay constant defined by the user before calculating the index.
Therefore, lower $\gamma$ indicates a stronger time decay. This approach was inspired by \cite{baruah:_dynamically_evolving_clustering_for_data_streams}.

\subsection{Normalization of the time weighting, linkage reliability and $\beta$ term}\label{app:normalization}

To make sure that the different time weightings and reliability values are comparable, we can calculate the normalized time-weighting values $\tilde{\alpha}_{mi}$ and $\tilde{\alpha}_{sj}$, normalized reliability values $\tilde{r}_{mi}$ and $\tilde{r}_{sj}$ as well as normalized $\beta$ term values $\tilde{\beta}_{mi}$ and $\tilde{\beta}_{sj}$ as follows:

\begin{equation}\label{eq:normalized_time_weighting_a_2}
	\tilde{\alpha}_{mi}=\frac{\alpha_{mi}}{\sum_{i \in D}^{} \alpha_{mi} + \sum_{k \in ch_{q}}^{} \sum_{j \in k}\alpha_{sj}+\sum_{k \in ch_{q}}^{} \alpha_{xk}}
\end{equation}

and

\begin{equation}\label{eq:normalized_time_weighting_b_2}
	\tilde{\alpha}_{sj}=\frac{\alpha_{sj}}{\sum_{i \in D}^{} \alpha_{mi} + \sum_{k \in ch_{q}}^{} \sum_{j \in k}\alpha_{sj}+\sum_{k \in ch_{q}}^{} \alpha_{xk}},
\end{equation}

where $\alpha$ is the raw non-normalized time-weighting that we defined in Eq.~\ref{eq:alpha}. Additionally, the normalized reliability values and normalized $\beta$ term values are calculated identically, by replacing the $\alpha$ terms in equations with $r$ or $\beta$, respectively. As a result of this normalization, the sum of all normalized terms is equal to one.

\end{appendices}

\end{document}